\documentclass{aa}

\usepackage{graphics, graphicx}
\usepackage{subcaption}
\usepackage{booktabs}
\usepackage{multirow}
\usepackage{tabularx}
\usepackage{ragged2e}
\newcolumntype{C}{>{\Centering}X}
\newcolumntype{M}[1]{>{\centering\arraybackslash}m{#1}}
\usepackage{adjustbox}
\usepackage{booktabs}
\usepackage[dvipsnames,table,xcdraw]{xcolor}
\usepackage[labelfont=bf]{caption}
\usepackage{pdfpages}
\usepackage{soul}
\usepackage{txfonts}
\usepackage[colorlinks=true, urlcolor=blue, linkcolor=blue, citecolor=blue, backref=page]{hyperref}
\usepackage{comment}
\usepackage{enumitem}
\usepackage[capitalise]{cleveref}
\crefname{section}{Sect.}{Sect.}
\newcommand{\HI}{\ion{H}{I}}
\newcommand{\Halpha}{\mathrm{H\alpha}}
\newcommand{\Hbeta}{\mathrm{H\beta}}
\usepackage{float}
\usepackage{oplotsymbl}
\defcitealias{Espada2026}{Paper I}
\usepackage[colorinlistoftodos]{todonotes}
\begin{document}
   \title{ALMA CO-CAVITY}
   \subtitle{II. Resolved Scaling Relations in Void Galaxies}

    \author{S. B. De Daniloff 
        \inst{\ref{inst:iram},\ref{inst:ugr}} \corrauth{sbonnal@iram.es}
    \and
        D. Espada \inst{\ref{inst:ugr},\ref{inst:ic1}} \email{despada@ugr.es}
    \and
        S. Duarte Puertas \inst{\ref{inst:ugr},\ref{inst:ic1}} \email{sduarte@ugr.es}
    \and
        Y. K. Gonz\'alez-Koda \inst{\ref{inst:ugr}} \email{yllari@ugr.es}
    \and
        G. Torres-R\'ios \inst{\ref{inst:ugr}} \email{gloriatr@ugr.es}
    \and
        S. F. S\'anchez \inst{\ref{inst:unam},\ref{inst:iac}} \email{sfsanchez@astro.unam.mx}
    \and
        R. Garc\'ia-Benito\inst{\ref{inst:iaa}} \email{rgb@iaa.es}
    \and
        S. Verley \inst{\ref{inst:ugr},\ref{inst:ic1}} \email{simon@ugr.es}
    \and
        I. P\'erez \inst{\ref{inst:ugr},\ref{inst:ic1}} \email{isa@ugr.es}
    \and 
        M. Argudo-Fern\'andez \inst{\ref{inst:ugr},\ref{inst:ic1}}
    \and 
        A. Bongiovanni \inst{\ref{inst:iram},\ref{inst:AsocAstro}} \email{bongio@iram.es}
    \and
        M. S\'anchez-Portal \inst{\ref{inst:iram},\ref{inst:AsocAstro}} \email{msanchez@iram.es}
    \and 
        U. Lisenfeld \inst{\ref{inst:ugr},\ref{inst:ic1}} \email{ute@ugr.es}
    \and
        M. I. Rodr\'iguez\inst{\ref{inst:iram}} \email{mrodriguez@iram.es}
    \and
        A. Conrado\inst{\ref{inst:iaa}} \email{aconrado@iaa.es}
    \and 
        A. Jim\'enez \inst{\ref{inst:ugr}} \email{andonijimenez@ugr.es}
    \and
        B. Bidaran \inst{\ref{inst:ugr}}
    \and
        L. S\'anchez-Menguiano \inst{\ref{inst:ugr},\ref{inst:ic1}} \email{lsanchezm@ugr.es}
    \and 
        T. Ruiz-Lara \inst{\ref{inst:ugr},\ref{inst:ic1}} \email{ruizlara@ugr.es}
    \and 
        A. Zurita \inst{\ref{inst:ugr},\ref{inst:ic1}} \email{azurita@ugr.es}
    \and
        P. V\'asquez-Bustos \inst{\ref{inst:ugr}} \email{paulovb@ugr.es}
    \and
        M. Alc\'azar-Laynez \inst{\ref{inst:ugr}} \email{manalclay@ugr.es}
    \and
        P. Villalba-Gonz\'alez \inst{\ref{inst:ubc}} \email{pedrovg@phas.ubc.ca}
    \and
        E. Florido \inst{\ref{inst:ugr},\ref{inst:ic1}} \email{estrella@ugr.es}
    \and
        R. E. Miura \inst{\ref{inst:ugr}} \email{riemiura@gmail.com}
   }

    \institute{Institut de Radioastronomie Millim\'etrique (IRAM), Av. Divina Pastora 7, N\'ucleo Central, E-18012, Granada, Spain \label{inst:iram}
        \and
            Dpto. de F\'isica Te\'orica y del Cosmos, Edificio Mecenas, Campus de Fuentenueva, Universidad de Granada, E-18071, Granada, Spain \label{inst:ugr}
        \and
            Instituto Carlos I de F\'isica Te\'orica y Computacional, Universidad de Granada, E-18071, Granada, Spain \label{inst:ic1}
        \and  
            Instituto de Astronom\'ia, Universidad Nacional Auton\'oma de M\'exico, A.P. 106, Ensenada 22800, BC, Mexico \label{inst:unam}
        \and
            Instituto de Astrof\'isica de Canarias, La Laguna, E-38200, Tenerife, Spain \label{inst:iac}
        \and 
            Instituto de Astrof\'isica de Andaluc\'ia (CSIC), PO Box 3004, E-18008, Granada, Spain \label{inst:iaa}
        \and
            Asociación Astrofísica para la Promoción de la Investigación, Instrumentación y su Desarrollo, ASPID, La Laguna, E-38205, Tenerife, Spain \label{inst:AsocAstro}
        \and 
            Department of Physics and Astronomy, University of British Columbia, Vancouver, BC, Canada \label{inst:ubc}
        }

    \date{}

    \abstract
    {
    Scaling relations involving star formation rates (SFRs), molecular gas mass, and stellar mass are key to understand galaxy evolution, and have previously been explored at resolved scales. However, they have not been examined with particular emphasis on the large-scale environments (LSEs). In this work, we study the resolved Schmidt-Kennicutt relation (rSK), molecular gas main sequence (rMGMS) and star-forming main sequence (rSFMS) from a sample of 41 void galaxies (VGs) residing in the least dense regions of the Universe. Using high-resolution interferometric CO(1--0) data and optical IFU data from the ALMA CO-CAVITY and CAVITY surveys at scales of $2\farcs5$ (0.8--2.1\,kpc), we study these relations for the full sample as well as for individual galaxies in voids. We fit the relations, finding a similar parametrisation as that used for galaxies from all LSEs. However, the rMGMS is the tightest of the three relations ($\sigma_\mathrm{rMGMS}=0.16\ \mathrm{dex}$, $\sigma_\mathrm{rSK}=0.21\ \mathrm{dex}$, and $\sigma_\mathrm{rSFMS}=0.24\ \mathrm{dex}$), unlike in other samples. We find that a large source of deviations in the relations comes from galaxy-to-galaxy variations. However, the rMGMS is less affected by these variations. It has been suggested that the rMGMS arises from the concentration of molecular gas within the gravitational potential set by the stellar content and dark matter. We hypothesise that deviations from the rMGMS trace changes in the gravitational potential occurring on longer time-scales, whereas deviations in the rSK and the rSFMS are driven by more rapid variations in the SFR. This distinction is particularly relevant for our sample because the 41 ALMA CO-CAVITY VGs are generally more isolated than galaxies in other LSEs, and therefore are less affected by events that can significantly alter the gas distribution or trigger SF on short time-scales. In this sense, the rMGMS is likely the most stable of these relations over time.
    }

    \titlerunning{ALMA CO-CAVITY II. Resolved Scaling Relations in Void Galaxies}
    \maketitle
    \nolinenumbers

\section{Introduction}\label{sec:introduction}
The properties of galaxies and the relations between them are key to understand galaxy evolution. In the past decades, many surveys have tried to establish the links between these properties in order to discover fundamental star-forming scaling relations (SRs; e.g. \citealt{Brinchmann2004,Catinella2010,Bothwell2013,Sanchez2021}) that could be related to the galaxy evolution. These SRs are maintained at global and local scales, which reveals fundamental processes within galaxies.

Some of the most meaningful SRs in the study of galaxy evolution are those which link their star formation rate (SFR) with molecular gas mass ($M_\mathrm{H_2}$), and stellar mass ($M_\star$). The Schmidt-Kennicutt relation (SK)~\citep{Schmidt1959,Kennicutt1998} connects the SFR with $M_\mathrm{H_2}$. This relation shows how the molecular gas acts as the fuel for star formation (SF). The molecular gas main sequence~\citep[MGMS;][]{Wong2013,Saintonge2016,Sanchez2018,Baker2023,Hagedorn2024} connects the galaxy $M_\mathrm{H_2}$ and $M_\star$. It has been less studied and is more difficult to interpret, although a possible interpretation is that the molecular gas concentrates within the gravitational potential set by the stellar and dark matter contents \citep{Lin2019,Baker2022}. Finally, the star-forming main sequence~\citep[SFMS;][]{Brinchmann2004,Salim2007} connects the galaxy SFR and $M_\star$, even at high redshift \citep[e.g.][]{Noeske2007,Speagle2014}. For SF galaxies, it relates the mass of stars that have already formed to the ongoing SF of the galaxy.

These global SRs, involving the integrated properties of the galaxies, have been extensively studied. However, the SRs at local scales of the disc, i.e. the so-called resolved SRs, are more challenging to obtain observationally. Following the pioneer works of \cite{Schmidt1959} and \cite{Kennicutt1998} on the global SK, several studies showed the relations between SFR surface densities ($\Sigma_\mathrm{SFR}$), $M_\mathrm{H_2}$ surface densities ($\Sigma_\mathrm{H_2}$), and atomic gas surface densities ($\Sigma_\mathrm{\HI}$) \citep[e.g.][]{Wong2002,Bigiel2008,Leroy2013}, namely the resolved Schmidt-Kennicutt relation (rSK). It is generally accepted that a tighter relation exists between $\Sigma_\mathrm{SFR}$ and $\Sigma_\mathrm{H_2}$ than with $\Sigma_\mathrm{\HI}$, which reflects the formation of stars in giant molecular clouds \citep{Bigiel2008}. The resolved molecular gas main sequence (rMGMS) has been more recently studied, with works from \cite{Lin2019}, \cite{Ellison2021a,Ellison2021b}, and \cite{Baker2022}, for example. These authors argue that this relation and the rSK are independent and fundamental. Meanwhile, the resolved star-forming main sequence (rSFMS) uses $\Sigma_\mathrm{SFR}$ and $M_\star$ surface densities ($\Sigma_\star$), and has been extensively studied at local scales \citep[e.g.][]{Sanchez2013,Wuyts2013,CanoDiaz2016,Ellison2020,Baker2022}. It is generally argued that this relation is close to linear or sub-linear\footnote{We define sub(super)-linear relations those with a slope in logarithmic scale lower (higher) than 1.} \citep{Sanchez2020}, with a higher dispersion than the other two \citep[e.g.][]{Lin2019}.

In the past decade, several surveys were carried out to methodically study the relations between these quantities. EDGE-CALIFA~\citep{Bolatto2017,Wong2024} is the first survey to produce results using a large sample, with 126 galaxies randomly selected from CALIFA DR3 \citep{Sanchez2016} at low redshift (${0.005<z<0.032}$), and an angular resolution of $7\arcsec$. ALMaQUEST~\citep{Lin2020} provides data for 46 galaxies (later extended to 66 galaxies, with redshift ${0.018<z<0.133}$) selected from the MaNGA sample and observed with the Atacama Large Sub/Millimeter Array (ALMA), with an angular resolution of $2\farcs5$ to study the SRs at kiloparsec scales. With closer galaxies ($d<27$\,Mpc), PHANGS~\citep{Leroy2021,Lee2022,Emsellem2022} has also been carrying out high-resolution observations at parcsec scales \citep[e.g.][]{Pessa2021,Neumann2025}. Other studies such as iEDGE~\citep{Colombo2025a} and SALVAGE~\citep{Wilkinson2026} combine multi-wavelength data from several datasets and with different resolutions to increment the number of galaxies at the cost of the resolution. All these surveys complement each other, allowing to trace SRs at different scales.

When studying the resolved SRs, the question arose as to which ones are more fundamental when tracing SF in galaxies. In \cite{Lin2019}, \cite{Ellison2021a}, and \cite{Baker2022}, the authors analysed the correlations and scatters of the relations within the ALMaQUEST sample, the galaxy-to-galaxy variability, and used techniques such as random forest regression analysis. They concluded that the rSK and rMGMS are the tightest and most correlated relations, that the rSFMS is derived from these two relations, and that $\Sigma_\mathrm{H_2}$ is the most important parameter driving $\Sigma_\mathrm{SFR}$. \cite{Wong2024} found that the correlation of the rSK is higher than that of the other two relations in the EDGE-CALIFA sample. Furthermore, \cite{Lin2019} and \cite{Sanchez2021} argue that the three quantities are following a linear relation in the phase-space diagram. Using a sample of 643 galaxies (iEDGE), \cite{Colombo2025b} show that a three-dimensional relation between the $M_\star$, $M_\mathrm{H_2}$ and SFR only exists for purely star-forming galaxies (while the relation exists, it is weaker for other types of galaxies). Despite all the achieved knowledge on these SRs, their connection to the environment of the galaxies has not yet been explored, especially at Mpc scales (large-scale environment; LSE).

Galaxies in the universe are distributed in large-scale structures. They inhabit clusters, which are nodes of high number density of galaxies, interconnected by filaments, walls, and under-dense regions called voids. Voids are particularly interesting since they allow for the study of galactic evolution with the most reduced impact of environmental processes coming from high-density environments. Although galaxies inhabiting voids, i.e. void galaxies (VGs), represent 10\% of the total count in the Universe, voids occupy 70\% of the volume of the universe \citep{Pan2012,Cautun2014}. Because they evolve in the most underdense environments, VGs are largely free from the external mechanisms that can influence galaxy properties in denser regions, making them valuable benchmarks for establishing canonical SRs, either resolved or the integrated ones.

The ALMA CO-CAVITY project~(\citealt{Espada2026}, hereafter \citetalias{Espada2026}) was specifically designed to study molecular gas properties in VGs, and investigate their relation with the SF and stellar content of galaxies. Using high-resolution observations from ALMA, this survey of 41 galaxies represents the first resolved-CO survey of VGs at kiloparsec scales. These data are combined with optical integral-field spectroscopy (IFS) data, which allow us to obtain resolved maps of $M_\star$ and SFR. \citetalias{Espada2026} presented the global SK, the MGMS and the SFMS of VGs, showing that the MGMS appears to be the tightest of the three relations and that there are no relevant differences in the global SRs of VGs and other samples of galaxies including denser LSEs.

In this paper we present the resolved SRs obtained for the first time at kiloparsec scales on a representative sample of VGs from the ALMA CO-CAVITY survey. In \cref{sec:methodology} we present the data and the quantities used in this work. In \cref{sec:full_sample_rSRs} we present the relations obtained from the full sample of pixels. In \cref{sec:individual_rSRs} we show the relations obtained for each galaxy individually and create a classification of galaxies based on their SRs. In \cref{sec:measuring_offsets_rSRs} we quantify the offsets in each relation and investigate their origin. In \cref{sec:discussion} we discuss the results to understand the role of the LSE on the variations of SRs. Finally, in \cref{sec:conclusions} we summarise the results and conclusions of this study.

In this work, we assume a $\Lambda$CDM cosmology with ${\mathrm{H_{0}}=70\ \rm km\ s^{-1}\ Mpc^{-1}}$, ${\mathrm{T_{CMB}}=2.725\ \rm K}$, and ${\Omega_{\mathrm{m},0}=0.3}$, and a Chabrier initial mass function \citep[IMF;][]{Chabrier_IMF}.

\section{Methodology}\label{sec:methodology}
The data of this work come from the ALMA CO-CAVITY survey~\citepalias{Espada2026}. The 41 VGs selected are star-forming galaxies with stellar masses over $10^{9.0}\ \mathrm{M_\odot}$, O3N2 \citep{pettini_oiiinii_2004} gas abundances ${\rm 12+\log(O/H) > 8.4}$, and optical diameters over 20\arcsec. This is a companion project of the Calar Alto Void Integral-field Treasury surveY (CAVITY; \citealt{Perez2024})\footnote{\url{https://cavity.caha.es/}}, a legacy survey of the Centro Astronómico Hispano en Andalucía (CAHA) that specifically targets IFU observations (using the Potsdam Multi Aperture Spectrophotometer fiber PAcK, PPak) of VGs, and tries to understand how their formation and evolution are driven by the low-density environment of voids. The ALMA CO-CAVITY survey uses IFU data cubes of 30 galaxies from CAVITY observations and of 11 galaxies from the MaNGA survey \citep{MaNGADR17}. Detailed information on the ALMA and IFU data can be found in the respective presentation papers. In the following subsections we focus on describing how the surface densities are derived.

\subsection{Derivation of $\Sigma_\mathrm{H_2}$}\label{sec:derivation_SigmaH2}
The $\Sigma_\mathrm{H_2}$ map is calculated from the $I_\mathrm{CO}$ moment-0 map of the $\rm ^{12}$CO(1--0) emission line. We refer the reader to \citetalias{Espada2026} for more information about the data products obtained using the CASA package \citep{CASA_Software} and ALMA pipeline, including data cubes and moment maps, as well as 3D masks to identify the CO(1--0) emission using the \texttt{SoFiA 2} pipeline. We used the masked CO(1--0) moment-0 maps, which focuses on the emission regions (reducing the effect of noise), and the global metallicity-dependent conversion factor $\alpha_\mathrm{CO}$ (\citetalias{Espada2026}, see also \citealt{Rodriguez2024}), that includes the correction for the presence of helium and heavier elements. In \cref{sec:app_variable_local_alphaCO} we demonstrate that using pixel-based instead of global values of $\alpha_\mathrm{CO}$ for each galaxy does not significantly change the resolved SRs and conclusions of this work.

For each galaxy we estimated a local uncertainty for the velocity-integrated line intensity, $\Delta I_\mathrm{CO}=\sigma\sqrt{\delta v \Delta V}$, where $\sigma$ is the RMS noise in units of $\rm Jy\ beam^{-1}$ in line-free channels and with a channel width of $\delta v\approx 10\ \mathrm{km\ s^{-1}}$ (values are available in Table A.2 of \citetalias{Espada2026}), and $\Delta V$ is the width of the CO(1--0) emission line in each spaxel. We then converted this intensity uncertainty into $\Sigma_\mathrm{H_2}$ uncertainty (see Subsect. 4.1 of \citetalias{Espada2026}).

We also estimated an upper limit for $\Sigma_\mathrm{H_2}$ ($\Sigma_\mathrm{H_2,\ u.\ l.}$). Following \cite{Pan2024}, we calculated the upper limit of $I_\mathrm{CO}$ to recover the emission for each spaxel as ${I_\mathrm{CO,\ u.\ l.}=3\sigma\sqrt{\delta v \Delta V_\mathrm{cloud}}}$, where ${\Delta V_\mathrm{cloud}=40\ \mathrm{km\ s^{-1}}}$ is a typical line width for a bulk of molecular clouds observed at kiloparsec resolution. Spaxels not included in the CO(1--0) mask were assigned the value of $\Sigma_\mathrm{H_2,\ u.\ l.}$ and identified as a non-detection.

\subsection{Derivation of $\Sigma_\mathrm{SFR}$ and $\Sigma_\star$}
The optical datacubes of the observed galaxies were analysed with the \texttt{pyPipe3D} pipeline~\citep{pyPipe3D2022} to obtain the spatially resolved gas and stellar population properties of the VGs, using the \texttt{MaStar\_sLOG} stellar library~\citep{Sanchez2022}. From the products of the pipeline \citep[presented in][]{CAVITY_pyPipe3D}, we used the spectral emission line fluxes, the $\Sigma_\star$ maps derived from spectral fitting of single stellar populations, the stellar dust extinction maps ($A_\mathrm{V, \star}$), and continuum dezonification maps (\texttt{DZ}), that are used to recover the spatial information after the Voronoi binning \citep{CidFernandes2013}\footnote{Since Voronoi binning combines the flux of spaxels within a zone, the spectral synthesis is performed once per zone, so all spaxels within a zone share the same fitted stellar population properties (age, metallicity, mass-to-light ratio, extinction). For extensive quantities like $\Sigma_\star$, spatial resolution is recovered via dezonification: each spaxel's $\Sigma_\star$ is obtained by scaling the zone's value by the spaxel's fractional contribution to the zone's total flux.}. From the value-added catalogue of integrated properties, we used the angular distance ($D_\mathrm{A}$, derived as $D_\mathrm{A}=D_\mathrm{L}/(1+z)^2$, where $D_\mathrm{L}$ is the luminosity distance and $z$ the redshift, extracted from SDSS DR7 and available in Table 1 of \citetalias{Espada2026}) to convert the pixel units into physical distances. For the galaxies included in the MaNGA sample, we directly used the \texttt{pyPipe3D} data products and catalogues of the MaNGA DR17 SDSS-IV\footnote{\url{https://www.sdss4.org/dr17/manga/manga-data/manga-pipe3d-value-added-catalog/}}, whose version is similar to that of the products obtained for the PPak observations. The rest of properties (celestial coordinates, $z$, $M_\star$, diameter $d_{25}$ at a 25 mag isophote, $\mathrm{12+\log(O/H)}$, position angle, inclination, and radius $r_{50}$ containing 50\% of the Petrosian flux in the SDSS $r$-band) are provided by the CAVITY database~\citep{Perez2024,Garcia_Benito2024}. We converted the products from \texttt{pyPipe3D}, that were calculated using the Salpeter IMF, to the Chabrier IMF by multiplying $M_\star$ by a factor of 0.61 and SFRs by 0.63~\citep{Madau_Dickinson}.

To compute the $\Sigma_\mathrm{SFR}$ maps, the SFR was calculated pixel-by-pixel using the $\Halpha$ flux corrected for extinction, and following the estimator from \cite{KennicuttEvans2012}, adapted to the Chabrier IMF: $\mathrm{SFR}\left(\mathrm{M_\odot \; yr^{-1}}\right)_\mathrm{\Halpha} = 5.04 \times 10^{-42}\ \left[\frac{L\left(\Halpha\right)}{\mathrm{erg\; s^{-1}}}\right]$, where $L(\Halpha)$ is the luminosity of $\Halpha$ emission line in $\mathrm{erg\; s^{-1}}$, calculated using its flux ($L = 4\pi D_\mathrm{L}^2 F$, where $D_\mathrm{L}$ is the luminosity distance in $\mathrm{cm}$, available in Table 1 in \citetalias{Espada2026}, and $F$ is the flux, originally provided by \texttt{pyPipe3D} in units of $10^{-16}\ \mathrm{erg\ s^{-1}\ cm^{-2}}$).

Furthermore, several corrections and conversions were applied to each map before they could be used in our analysis:
\begin{itemize}[leftmargin=0pt, itemsep=0pt, topsep=0pt, label={}]
    \item $-$ Galactic foreground extinction: the CAVITY datacubes were processed and corrected for Galactic extinction \citep{Garcia_Benito2024} using the CCM89 \citep{CCM89} empirical function and maps from \cite{Schlegel1998}. Unlike the CAVITY datacubes, MaNGA data are not originally corrected for Galactic foreground extinction\footnote{\url{https://www.sdss4.org/dr17/manga/manga-analysis-pipeline/}}. We apply this correction using the total extinction map provided by \texttt{pyPipe3D}, which includes the Galactic foreground extinction as well as stellar and nebular extinctions (see next).

    \item $-$ Stellar dust extinction correction (only for $\Sigma_\star$): ${\log_{10}(\Sigma_\mathrm{\star,\,\, corr}) = \log_{10}(\Sigma_\star) + 0.4A_\mathrm{V, \star}}$.
    
    \item $-$ Nebular dust extinction correction (only for $\Sigma_\mathrm{SFR}$): the flux of the H$\alpha$ emission line is corrected for dust extinction using the Balmer decrement and adopting an intrinsic ratio ${(F_\Halpha/F_\Hbeta)_{\mathrm{int}} = 2.86}$, corresponding to a temperature $T=10^4$ K and an electron density $n_\mathrm{e} = 10^2\ \mathrm{cm^{-3}}$ for Case B recombination \citep{Osterbrock1989}. We calculate the reddening coefficient, $C(\Hbeta)$, following the CCM89 empirical function with $R_\mathrm{V}=3.1$. In cases where the observed Balmer decrement is below 2.86, we do not apply the dust extinction.
    
    \item $-$ Dezonification (only for $\Sigma_\star$): ${\log_{10}(\Sigma_\mathrm{\star,\ corr}) = \log_{10}(\Sigma_\star) + \log_{10}(\texttt{DZ})}$, where $\texttt{DZ}$ is the continuum dezonification map provided by \texttt{pyPipe3D}.
    
    \item $-$ Inclination correction: ${\log_{10}(\Sigma_\mathrm{corr}) = \log_{10}(\Sigma) + \log_{10}\left[\cos(i)\right]}$, where $i$ is the galaxy inclination angle. This correction is applied to all the surface density maps, including $\Sigma_\mathrm{H_2}$.
\end{itemize}

\subsection{Data homogenisation}
ALMA maps have resolutions of $\sim0\farcs8-1\farcs9$ and pixel sizes of $\sim0\farcs18$ (see Table A.2 in \citetalias{Espada2026} for details on the data products). Following the procedure detailed in \cref{sec:derivation_SigmaH2}, we obtained $\Sigma_\mathrm{H_2}$ maps at native spatial resolution. Since PPak and MaNGA optical maps have nominal resolutions of $2\farcs5$, we smoothed ALMA {moment-0} maps to a Gaussian circular kernel of $2\farcs5$, corresponding to linear resolutions of 0.8--2.1\,kpc. Additionally, we regridded the smoothed ALMA maps to their corresponding optical maps as templates (pixel size $0\farcs5$ for MaNGA and $1\farcs0$ for PPak data) using the \texttt{imregrid} task from CASA. We show an example of how the data are modified in \cref{fig:smoothing_example}.

\subsection{Selection of detected pixels}\label{sec:selection_det_spax}
For $\Sigma_\mathrm{H_2}$, we consider `detected pixels' if ${I_\mathrm{CO} > 3\sigma\sqrt{\delta v \Delta V_\mathrm{cloud}}}$, for $\Sigma_\mathrm{SFR}$ if the $\Halpha$ and $\Hbeta$ emission lines S/N are above 3, and for $\Sigma_\star$ if the average continuum S/N is above 3.

Furthermore, to exclude from the analysis regions located outside the galaxy projected disc, we created elliptical de-projected masks, using their $d_{25}$ as the major axis, inclination and position angle (Table 1, \citetalias{Espada2026}). We obtain a total of 7\,937 pixels above the detection limits in all the three quantities, for a total of 35 VGs with CO detection.

\section{Resolved scaling relations for the full sample}\label{sec:full_sample_rSRs}
In this section we present the resolved SRs (rSK, rMGMS, and rSFMS) obtained when combining the pixels from the full ALMA CO-CAVITY sample and discuss them in the context of previous surveys. To complement this analysis, in \cref{sec:app_rSRs_23axes} we study them using pixels detected in only two of the three axes, and, in \cref{sec:app_survival_analysis}, including upper limits. The inclusion of these additional data does not significantly alter the relations and further supports our conclusions.

\begin{figure*}
    \centering
    \includegraphics[width=\textwidth]{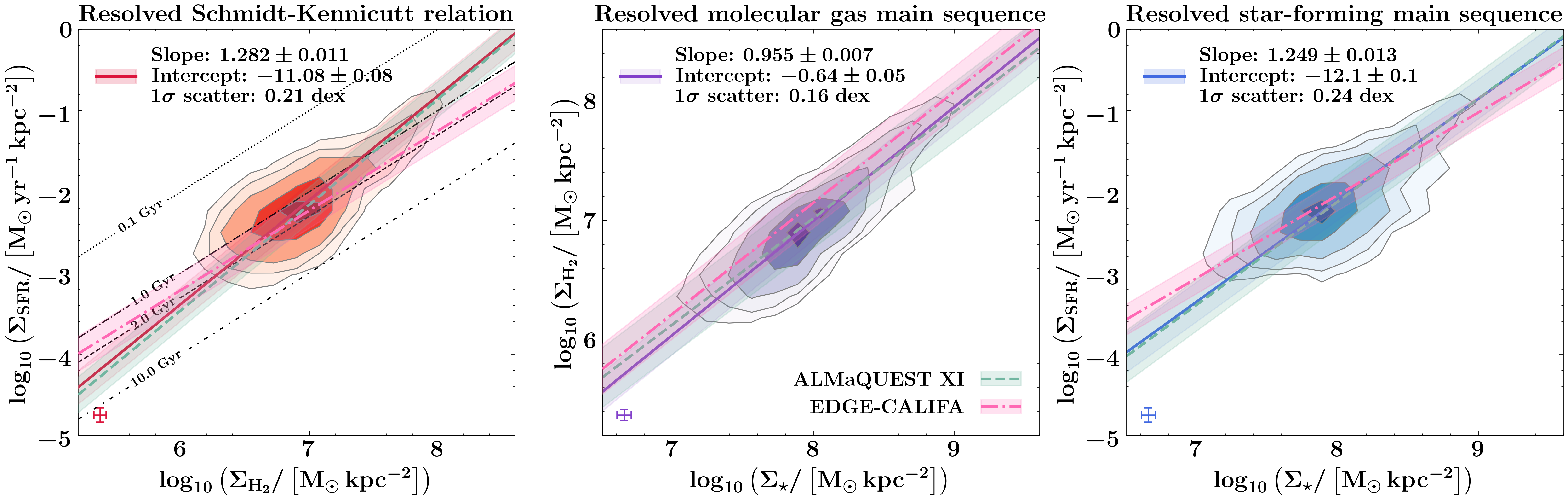}
    \caption{Resolved SRs for the 41 ALMA CO-CAVITY galaxies. We show the distribution of 7\,937 detected pixels with the filled contours: (left) in red for the rSK, (mid) in purple for the rMGMS, and (right) in blue for the rSFMS. In each panel, we show in the lower-left corner the median of the error bars. The orthogonal distance regression (ODR) fit results are shown in the upper-left legend of each panel and are represented by the solid lines, with the 1$\sigma$ ODR scatter depicted by the shadowed region around the regression. In the case of the rSK, we also show the equivalent slopes for depletion times at 0.1, 1.0, 2.0, and 10.0 Gyr, with black dotted and dashed lines. For comparison, we added the fits of ALMaQUEST XI~\citep{Ellison2024} full sample and EDGE-CALIFA~\citep{Sanchez2021} detected sample, with the green dashed and magenta dash-dotted lines, respectively. Note that the contours show levels of inclusion of pixels of 5, 10, 20, 50 and 80\%.}
    \label{fig:triple_general_plot_ODR}
\end{figure*}

In \cref{fig:triple_general_plot_ODR} we show the rSK (left), the rMGMS (mid), and the rSFMS (right), with all the detected pixels in the three variables. We fit each relation applying an orthogonal distance regression (ODR; solid lines) on the values without weighting by uncertainties. We indicate the parameters for the fit, including the 1$\sigma$ ODR scatter (calculated using the residuals in both axes), in \cref{tab:rSRs_all_fits_full_sample_surveys}. We also added the results of the fits obtained by ALMaQUEST XI full and control samples \citep{Ellison2024} and EDGE-CALIFA~\citep{Sanchez2021} surveys for comparison.

\begin{table}[h]
\caption{Linear-fit parameters to the resolved SRs for the samples of detected pixels of this work (ODR fits), ALMaQUEST XI~\citep{Ellison2024}, and EDGE-CALIFA~\citep{Sanchez2021}.}
\label{tab:rSRs_all_fits_full_sample_surveys}
\resizebox{\hsize}{!}{
    \begin{tabular}{cccc}
        \hline\hline
        \multicolumn{1}{c}{Survey} & Slope & Intercept & Scatter \\
                                   &       &           &  (dex)  \\  \hline
        \multicolumn{4}{c}{\textit{rSK}}                             \\ \hline
        This work    & $1.282 \pm 0.011$      & $-11.08 \pm 0.08$ & $0.21$    \\
        ALMaQUEST XI (FS) & $1.30 \pm 0.01$   & $-11.26 \pm 0.05$ & $0.22^\dagger$    \\
        ALMaQUEST XI (CS) & $1.35 \pm 0.01$   & $-11.63 \pm 0.08$ & $0.19^\dagger$    \\
        EDGE-CALIFA       & $0.98 \pm 0.14$   & $-9.01 \pm 0.14$  & $0.249$   \\ \hline
        \multicolumn{4}{c}{\textit{rMGMS}}                                    \\ \hline
        This work    & $0.955 \pm 0.007$      & $-0.64 \pm 0.05$ & $0.16^\dagger$     \\
        ALMaQUEST XI (FS) & $0.89 \pm 0.01$   & $-0.10 \pm 0.05$  & $0.25$     \\
        ALMaQUEST XI (CS) & $0.48 \pm 0.01$   & $3.25 \pm 0.06$   & $0.26$     \\
        EDGE-CALIFA       & $0.93 \pm 0.11$   & $-0.91 \pm 0.16$  & $0.218^\dagger$        \\ \hline
        \multicolumn{4}{c}{\textit{rSFMS}}                                    \\ \hline
        This work    & $1.249 \pm 0.013$      & $-12.70 \pm 0.11$ & $0.24$    \\
        ALMaQUEST XI (FS) & $1.27 \pm 0.01$   & $-12.28 \pm 0.08$ & $0.32$    \\
        ALMaQUEST XI (CS) & $0.65 \pm 0.01$   & $-7.26 \pm 0.08$  & $0.34$    \\
        EDGE-CALIFA       & $1.02 \pm 0.16$   & $-10.10 \pm 0.22$ & $0.266$   \\ \hline \hline
    \end{tabular}}
    \tablefoot{For each survey we converted the parameters to $\rm yr^{-1}$ and $\rm pc^{-2}$ units, and to Chabrier IMF. In the case of ALMaQUEST XI, we show the parameters for the fit applied to the full sample (FS), to the control sample (CS), and to the central starburst sample (SB). 
    
    $^\dagger$ In the last column, the dagger symbols show the relation with the lowest scatter for each survey/sample. Note that the fits from EDGE-CALIFA are not obtained using the ODR fitting procedure used in this work and ALMaQUEST XI.}
\end{table}

The resolved SRs of the ALMA CO-CAVITY sample show strong correlations with scatters between $0.16-0.24$\,dex. The slopes of the rSK and the rSFMS are super-linear, whereas the slope of the rMGMS is sub- or close to a linear relation. The obtained parameters at galactic scales are consistent with the global fits of VGs in \citetalias{Espada2026}.

The data points of the rSK concentrate close to a molecular gas depletion times ($\tau_\mathrm{dep}$) around $1-2$\,Gyr, which is typically found in other surveys \citep[e.g.][]{Bigiel2008,Verley2010}. We find for our sample of VGs a super-linear slope, similar to the works of \cite{Ellison2021a} and \cite{Ellison2024} (except starbursts), but in contrast to \citet{Bigiel2008} and \cite{Leroy2013}, for a smaller sample. In these cases the samples were selected regardless of the environment, and the ranges of applicability of the fits were similar to ours. A possible cause for the higher slopes is that here we are using low-J transitions of CO that trace low-density molecular gas, whereas other works employ dense gas tracers such as HCN, which are more appropriate to trace the mass of the gas used to form new stars, inducing linearity in its relation with the SFR \citep{Wu2005}.

The fits for all SRs for the VGs are in agreement with those of ALMaQUEST XI~\citep{Ellison2024}, with similar slopes and intercepts. However, one notable difference is the scatter obtained for each fit: the rMGMS is the tightest of the relations for VGs, with a scatter 0.05\,dex lower than that of the rSK and 0.08\,dex lower than that of the rSFMS, whereas for ALMaQUEST full and control samples, the rSK is the tightest relation (see \cref{tab:rSRs_all_fits_full_sample_surveys}). When we compare their Figs. 1 to 3 with our \cref{fig:triple_general_plot_ODR}, we note that the values in the ALMaQUEST XI sample are distributed more evenly across the entire range. In our sample, however, the surface densities are concentrated at $\log_{10}\left(\Sigma_\mathrm{H_2}/[\mathrm{M_\odot\ kpc^{-2}}]\right)\sim7.0$ and $\log_{10}\left(\Sigma_\mathrm{SFR}/[\mathrm{M_\odot\ yr^{-1} kpc^{-2}}]\right)\sim-2.2$, resulting in higher scatter in the fit. In contrast, the gas and stellar mass distribution underlying the rMGMS are more similar when comparing ALMaQUEST with CAVITY. With respect to EDGE-CALIFA our fits are overall steeper, except in the case of the rMGMS fit, whose slope is similar to that of ALMA CO-CAVITY. As mentioned in \citetalias{Espada2026}, EDGE-CALIFA galaxies are closer, from 23 to 130\,Mpc vs 70 to 180\,Mpc for ALMA CO-CAVITY galaxies, and the typical angular resolution of the data is $\sim4\farcs5\approx1.4$\,kpc, so it is likely that this discrepancy in the fits arises from the differences between surveys. Nevertheless, we observe that the scatter of the fits for the EDGE-CALIFA sample follows the same order as ours: the rMGMS is the tightest, followed by the rSK, and the rSFMS has the largest scatter. However, the differences between the scatter values of the ALMA CO-CAVITY sample are larger, at 0.08 dex between the rMGMS and rSFMS, compared to 0.04 dex for EDGE-CALIFA, and we also note that the comparison is risky as \cite{Sanchez2021} does not use the same fitting procedure to estimate the scatter.

We also note that by removing 14 galaxies with a potential companion within 1\,Mpc projected radius and $\pm500\ \mathrm{km\ s^{-1}}$ in line-of-sight velocity (following Table 2 of \citetalias{Espada2026}), the scatter of the fit to the resolved SRs remain unchanged (from 0.16\,dex to 0.17\,dex for the rMGMS, whereas it does not change for the rSK and rSFMS). The local environment of the studied galaxies is not significantly affecting the resolved SRs for our sample.

The rSFMS presents the largest scatter (0.24\,dex) of the three relations, in agreement with previous studies \citep[e.g.][]{Pan2024,Sanchez2021}. This relation presents the largest difference with respect to EDGE-CALIFA, whose fit to the rSFMS is much shallower than ours.

Finally, to evaluate the robustness of our result and the comparison between the 1$\sigma$ ODR scatters we make between resolved SRs, we applied a bootstrap random sampling on the data. With a total of 10\,000 bootstrap resamples, we obtained a distribution of scatter values for each relation, and we observed if the mean and standard deviation of each distribution coincide with the values presented in \cref{tab:rSRs_all_fits_full_sample_surveys}. The mean rSK ODR scatter is 0.2065\,dex, with a standard deviation of 0.0018\,dex, the mean rMGMS ODR scatter is 0.1614\,dex with a standard deviation of 0.0017\,dex, and the mean rSFMS ODR scatter is 0.2443\,dex with a standard deviation of 0.0022\,dex. These results coincide with that presented above, and confirm that the differences between the ODR scatters we observe are statistically significant.

\section{Resolved scaling relations for individual VGs}\label{sec:individual_rSRs}
The study of resolved SRs presents significant challenges. For example, the slope of these relations depends on the choice of the SFR indicator and the spatial resolution of the data \citep[e.g.][]{Hani2020,Chevance2020}, while the associated scatter is strongly influenced by sample selection and pixel size \citep{Pessa2022}. In addition, individual galaxies follow distinct evolutionary pathways and may undergo events such as mergers or quenching, which can further impact their SRs \citep[e.g.][]{Thorp2022,Dou2021}. For these reasons, it is relevant to show the SRs obtained for each of the 41 galaxies individually, which is the focus of the next subsections.

\subsection{The resolved Schmidt-Kennicutt relation}\label{sec:rSK_individual}
\begin{figure*}[ht]
    \centering
    \sidecaption
    \includegraphics[width=0.7\textwidth]{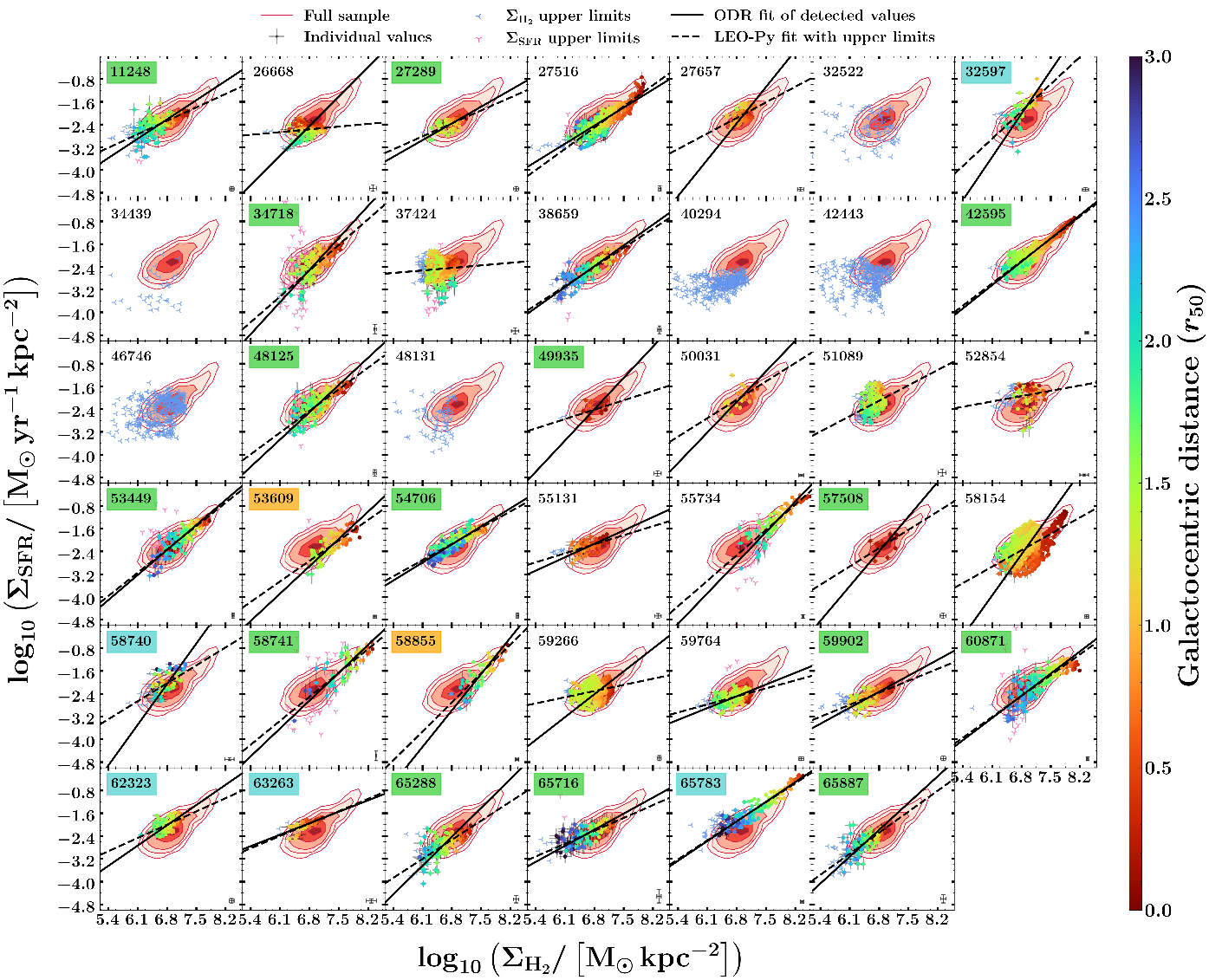}
    \caption{Individual resolved Schmidt-Kennicutt relations for each of the ALMA CO-CAVITY VGs. The red contours show the distribution of the full sample of detected pixels. Coloured data points are the individual detected values for each galaxy, with the colour indicating the galactocentric distance normalised by the $r_{50}$ represented by the vertical colour bar on the right side of the figure. In the upper-left corner of each panel, names in green are from galaxies whose rSK values do not deviate from the full-sample trend, in blue from those whose rSK values are above, and in orange from those below. The median error bars of the detected values are shown in the lower-right corner. We show the $\Sigma_\mathrm{H_2}$ upper limits with blue triangular markers, and the $\Sigma_\mathrm{SFR}$ upper limits with pink triangular markers. Non-detections in both variables are not shown. For the detected values, we perform an ODR fit without including upper limits nor errors, depicted with the black solid line when the fit was successful. All the values, including detected values with errors, and the upper limits, are fitted with \texttt{LEO-Py} pipeline, and the fit is depicted with a black dashed line. Galaxies without $\Sigma_\mathrm{H_2}$ detections do not have any linear fit.}
    \label{fig:rSK_all_galaxies}
\end{figure*}

The individual rSKs are displayed in \cref{fig:rSK_all_galaxies}, with the CAVITY ID given in the top left corner of each panel. In each panel, we show the distribution for the full sample of 7\,937 pixels (red contours), as well as upper limits. The data points are colour-coded by their deprojected galactocentric distance. The values were fitted with and without consideration of the upper limits (see \cref{sec:app_survival_analysis} for a description of the analysis using upper limits). In cases where the ODR fit provided a slope higher than 4.0, or a relative error higher than 60\% in any of the two parameters (slope and intercept), we considered the fit as failed. The resulting parameters for all the fits are available in \cref{sec:app_parameters_linear_fits}.

We observe significant variations between galaxies. Galaxies such as \texttt{11248}, \texttt{27289} or \texttt{38659} display similar trends as the full-sample distribution for all galaxies, that shows a major concentration at $\log_{10}(\Sigma_\mathrm{H_2}/[\mathrm{M_\odot\ kpc^{-2}}]) \in [6.5,\ 7.5]$ and $\log_{10}(\Sigma_\mathrm{SFR}/[\mathrm{M_\odot\ yr^{-1}\ kpc^{-2}}]) \in [-3.0,\ -1.5]$. Other cases like \texttt{32597}, \texttt{58740}, and \texttt{65783} display values above the full-sample distribution of points, demonstrating enhanced star-formation efficiency (SFE), whereas galaxies like \texttt{53609} and \texttt{58855} show values below the full-sample distribution, possibly related to a lower SFE. The slopes of the ODR fits range from $0.6$ to $2.2$, with a mean value of $1.29$ and a standard deviation of $0.43$, and the intercepts range from $-16.8$ to $-5.9$ with a mean value of $-11.10$ and a standard deviation of $3.0$. The large standard deviation shows the high variability between galaxies.

These differences arise from the intrinsic diversity found in our sample. In some cases, we discern several trends based on the position of the pixels within the plane of the galaxy. In the case of \texttt{58154}, we observe two distinct trends: one cluster following the full-sample distribution, and a second, straight line (in red), formed by data points from the central part of the galaxy, where a bar is visible (see Fig. 1 in \citetalias{Espada2026}). When looking at the individual surface density maps (available in \citetalias{Espada2026}), enhanced $\Sigma_\mathrm{SFR}$ and $\Sigma_\mathrm{H_2}$ are also visible in the spiral arms of the galaxy. \texttt{59266} is a galaxy that also displays several spiral arms, but its $\Sigma_\mathrm{SFR}$ is distributed more evenly across the disc (despite its arms being observed at a similar linear resolution as for \texttt{58154}). This results in values of $\log_{10}(\Sigma_\mathrm{H_2}/[\mathrm{M_\odot\ kpc^{-2}}])$ increasing up to $7.25$ towards the centre of the galaxy, while $\log_{10}(\Sigma_\mathrm{SFR}/[\mathrm{M_\odot\ yr^{-1}\ kpc^{-2}}])$ remains constant around $-2.5$. These two examples show that apart from the galaxy-to-galaxy differences visible when studying the resolved SRs, it is also relevant to analyse how these relations vary within galaxy substructures.

\subsection{The resolved molecular gas main sequence}\label{sec:rMGMS_individual}
\begin{figure*}[ht]
    \centering
    \sidecaption
    \includegraphics[width=0.7\textwidth]{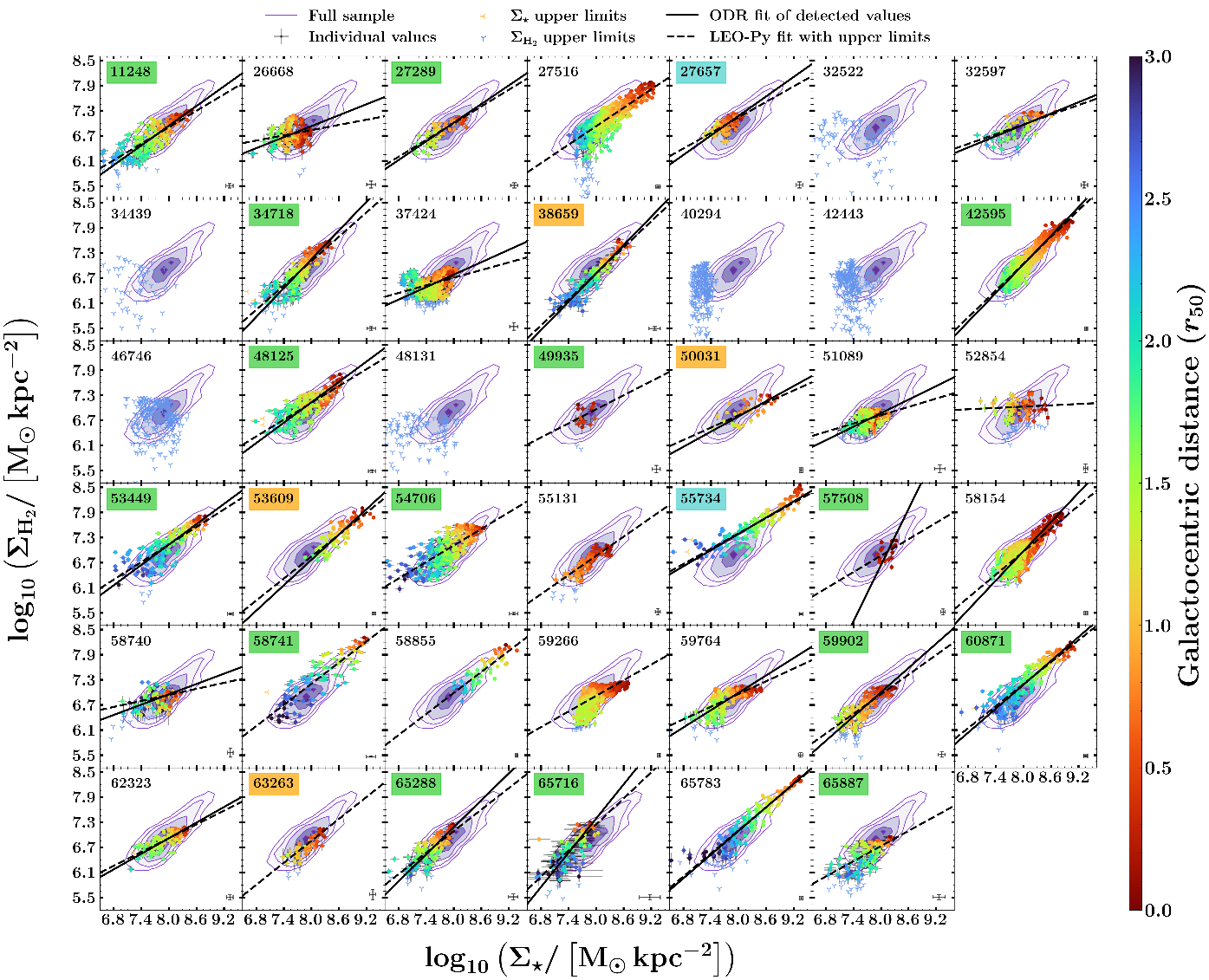}
    \caption{Resolved molecular gas main sequence for all ALMA CO-CAVITY VGs. The symbols and colours are equivalent to those of \cref{fig:rSK_all_galaxies}. $\Sigma_\star$ upper limits are depicted with yellow triangular markers.}
    \label{fig:rMGMS_all_galaxies}
\end{figure*}

In \cref{fig:rMGMS_all_galaxies} we present the rMGMS for the 41 VGs. There are fewer differences between galaxies than in the rSK. The mean value of the slopes is $0.88$ and the standard deviation is $0.39$ (full range from $0.4$ to $2.2$); and $-0.02$ and $3.11$ for the intercepts, respectively (full range from $-11.3$ to $3.7$). This is the relation with the smallest scatter (\cref{fig:triple_general_plot_ODR}), because there are less galaxy-to-galaxy variations and between substructures.

Despite the similarities between galaxies, we still observe some particular cases. \texttt{55734}, for instance, shows a relation 0.5\,dex above the full-sample distribution, whereas \texttt{38659} is displaying values slightly below. Galaxies \texttt{59266} and \texttt{59764} show a flattening of their relation at high $\Sigma_\star$, although other cases that appear to have similarly high values of $\Sigma_\star$ (${\log_{10}(\Sigma_\star/[\mathrm{M_\odot\ kpc^{-2}}]) > 8.5}$) such as \texttt{53609}, \texttt{55734}, \texttt{58154} or \texttt{58741} do not display such trend. Furthermore, both galaxies reaches the plateau at a similar value of ${\log_{10}(\Sigma_\mathrm{H_2} / [\mathrm{M_\odot\ kpc^{-2}}])\sim7.25}$. While the $\Sigma_\star$ distribution peaks at the centre of the galaxies, the $\Sigma_\mathrm{H_2}$ distribution is more evenly distributed across the discs, which causes the plateau in the rMGMS of these two galaxies. Finally, galaxies \texttt{26668}, \texttt{51089} and \texttt{52854} show little dynamic range in $\Sigma_\mathrm{H_2}$, which was also visible in their rSK.

\subsection{The resolved star-forming main sequence}\label{sec:rSFMS_individual}
\begin{figure*}[ht]
    \centering
    \sidecaption
    \includegraphics[width=0.7\textwidth]{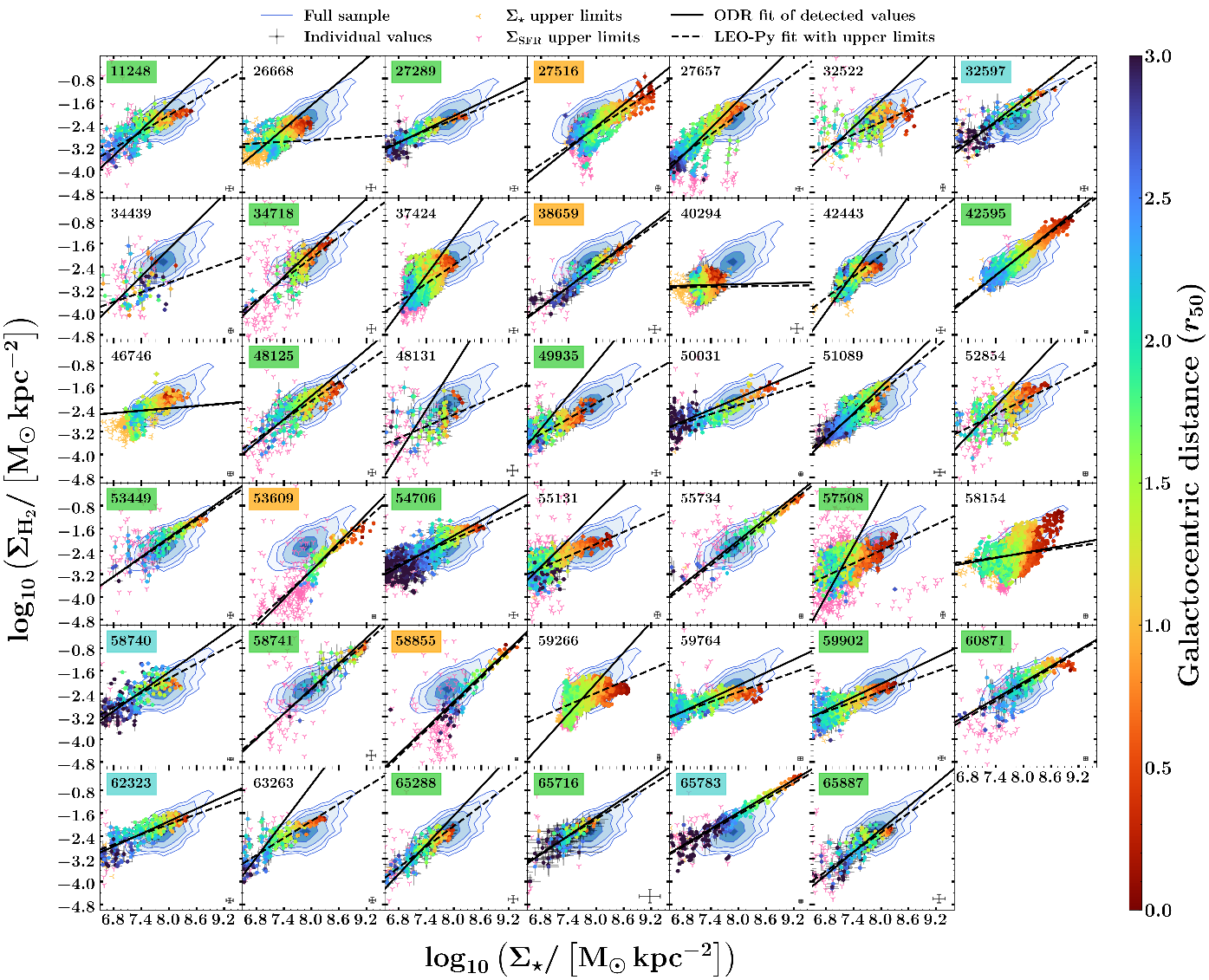}
    \caption{Resolved star-forming main sequence for all ALMA CO-CAVITY VGs. The symbols and colours are equivalent to those of \cref{fig:rSK_all_galaxies,fig:rMGMS_all_galaxies}.}
    \label{fig:rSFMS_all_galaxies}
\end{figure*}

In \cref{fig:rSFMS_all_galaxies} we show the rSFMS for the 41 VGs. The mean of the slopes is $1.36$ and the standard deviation is $0.60$ (full range from $0.04$ to $2.9$), and the mean of intercept is $-12.66$ with a standard deviation of $4.41$ (full range from $-24.0$ to $-3.4$).

We also observe, as for the previous resolved SRs, peculiar trends for some galaxies. The galaxies that differ the most from the others have lower $\Sigma_\mathrm{SFR}$ or higher $\Sigma_\star$: \texttt{27516}, \texttt{38659}, \texttt{53609}, and \texttt{58855} show distinctive trends below the full-sample distribution. On the other hand, others like \texttt{32597}, \texttt{58740} and \texttt{65783}, are displaying relations above the full-sample distribution.

\subsection{Galaxy classification based on their scaling relations}\label{sec:groups_rSRs}
To assess the heterogeneity of the sample, we visually group the galaxies in five categories based on their observed SRs with respect to the full sample: (i) following the same trend, (ii) different trend, (iii) individual relation following several trends, (iv) above, and (v) below. To help with this visual classification, we created an atlas with plots of the three resolved SRs, side by side, that is available electronically. In \cref{sec:app_categories_galaxies_rSRs} we present \cref{tab:categories_rSRs_ids} with the classification of all the galaxies, as well as the relations of representative cases for each category.

While galaxies with complex deviations of their relations across the disc would require individual studies at the scale of the disc, distributions above and below the full-sample trend can be more easily identified and studied. In \cref{fig:rSK_all_galaxies,fig:rMGMS_all_galaxies,fig:rSFMS_all_galaxies}, galaxies whose name has a cyan background colour are classified above the corresponding full-sample trend, while those with an orange background colour are classified below it. Focusing on these galaxies, we identify below the four groups that stand out for the trends in their three resolved SRs. We make use of properties presented in \citetalias{Espada2026}: the T-type from \cite{Dominguez_Sanchez2018}, which is a continuous morphological type associated with the classical Hubble sequence, that corresponds to early-type ($\mathrm{T-type}\leq0$, comprising elliptical and S0 galaxies) and spiral galaxies ($\mathrm{T-type}>0$, from Sa to Sm); the classification into SF (star-forming), TZ (transition zone), or SB (starburst) from \cite{Janowiecki2020}; the $M_\star$, the SFR and specific SFR (sSFR), and the metallicity (12+log(O/H); and the normalized void-centric distance, expressed as $r/R_\mathrm{eff}$, where $R_\mathrm{eff}$ is the effective radius of the void).

\begin{itemize}[leftmargin=0pt, itemsep=0pt, topsep=0pt, label={}]
    \item $-$ \underline{rSK and rSFMS above, rMGMS normal}: the data points of \texttt{32597}, \texttt{58740}, \texttt{62323}, \texttt{63263}, and \texttt{65783} are systematically above the full-sample rSK and rSFMS, but following the full-sample rMGMS trend. These galaxies present similar $\rm \log_{10}(sSFR/yr^{-1})\sim [-10.0, -9.4]$ (the standard deviation for the full sample is 0.36\,dex). As the sSFR is the SFR per unit of $M_\star$, it seems logical to observe similar trends on the relations involving $\Sigma_\mathrm{SFR}$ and $\Sigma_\star$ for these galaxies.
    
    \item $-$ \underline{rSK and rSFMS normal, rMGMS above}: \texttt{27657} and \texttt{55734} present typical trends with respect to the full-sample distribution in their rSK and rSFMS, but their rMGMS values are slightly higher. These galaxies are SB galaxies, but unlike the previous group, their global properties do not show clear similarities. For lower $\Sigma_\star$, their $\Sigma_\mathrm{H_2}$ values tend to be higher, whereas for higher values of $\Sigma_\star$, the values of $\Sigma_\mathrm{H_2}$ follow the full-sample trend.

    \item $-$ \underline{rSK and rSFMS below, rMGMS normal}: \texttt{53609} and \texttt{58855} form a third group, with rSK and rSFMS below the full-sample distribution, while the rMGMS follows the global trend. In this case, both galaxies are early-type (T-type of $-2.3$ and $-1.6$, respectively), they are massive galaxies with $\log_{10}(M_\star/\mathrm{M_\odot})\sim10.4$ and $10.2$, respectively, and they are both found at $0.6R_\mathrm{eff}$ of the centre of their void. Similarly to the first group, this finding could indicate that the rMGMS is less affected by the phenomena that is suppressing the SF in these galaxies.

    \item $-$ \underline{Non-detected galaxies}: the galaxies with no CO(1--0) detections are \texttt{32522}, \texttt{34439}, \texttt{40294}, \texttt{42443}, \texttt{46746}, and \texttt{48131}. Their $M_\star$ are $\sim 10^{9.2-9.4}\ \mathrm{M_\odot}$ and their T-type indicate they are late-type galaxies, that in some cases present spiral arms or a central bar. These galaxies have high $\alpha_\mathrm{CO}$ values, between 5.82 and 8.17 $\rm M_\odot\ (K\ km\ s^{-1}\ pc^2)^{-1}$ (while the mean value for our sample is 4.56 and the standard deviation 1.66), which is almost or more than twice the value for the Milky Way \citep[][and references therein]{Bolatto2013}, even though no molecular gas was detected using ALMA. We cannot study the similarities between resolved SRs as for the other groups, but we can still analyse their rSFMS. On one side, \texttt{32522}, \texttt{34439} and \texttt{48131} show dispersed values in the rSFMS. On the other side, \texttt{40294}, \texttt{42443} and \texttt{46746} show steep trends in the rSFMS. These three galaxies are classified as star-forming galaxies, with similar global values of SFR, $M_\star$, and $\rm 12+\log(O/H)$. Nevertheless, their rSFMS are different, with \texttt{40294} displaying clustered values around $\mathrm{10^{7.25}\ M_\odot\ kpc^{-2}}$ with almost no correlation between $\Sigma_\mathrm{SFR}$ and $\Sigma_\star$, whereas \texttt{46746} values are better distributed across the full-sample distribution, and \texttt{42443} values are found in between both states. These galaxies may present some bar features (especially \texttt{40294}), possibly concentrating and locally increasing $\Sigma_\star$ (\citealt{Seo2019}; see \citealt{JamesPercival2016,Neumann2024} and references therein for a review on how stellar population can be affected by bar features). This could cause an inside-out quenching of the SF \citep[e.g.][]{Scaloni2024,Veronese2025}, creating a decorrelation between $\Sigma_\star$ and $\Sigma_\mathrm{SFR}$ values, as we see it for \texttt{40294}. However, the bar should concentrate the molecular gas in the centre of the galaxy and increase its detectability, and this is not the case here.
\end{itemize}

\section{Offsets in the resolved scaling relations}\label{sec:measuring_offsets_rSRs}
As mentioned in \cref{sec:introduction}, the rSK and the rMGMS are considered to be the tightest and likely most independent relations \citep[e.g.][]{Lin2019,Ellison2021a}, while the rSFMS is a by-product of the previous two \citep[e.g.][]{Baker2022}. A signature of the potential fundamentality of each relation comes from their scatter \citep{Lin2020,Ellison2021a}. The ODR fits we applied in \cref{sec:full_sample_rSRs} show that the rMGMS is the tightest relation, partly due to its small galaxy-to-galaxy variations (see \cref{sec:rMGMS_individual}). In this section we explore if the galaxy-to-galaxy variations are effectively causing the scatter of the full-sample relations.

\begin{figure*}
    \centering
    \includegraphics[width=\textwidth]{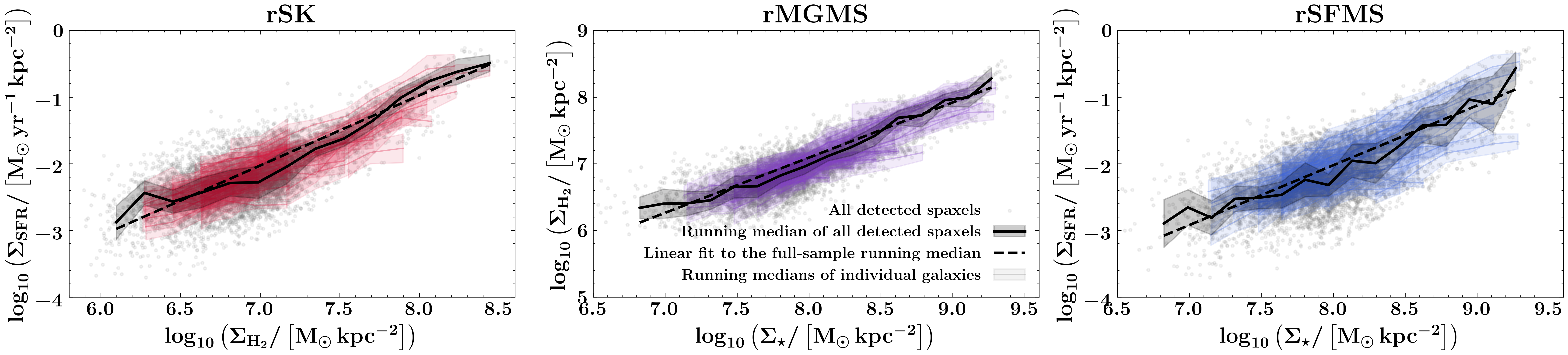}
    \caption{Running medians of resolved SRs for the rSK (left panel), rMGMS (mid panel) and rSFMS (right panel), for the 35 detected galaxies. In each panel, we show the individual detected pixels with grey data points, the running medians calculated for each galaxy individually with the coloured curves, and the one calculated for the full-sample distribution with the black curve. We apply a linear fit to the full-sample running medians depicted with the dashed black line. To calculate the running medians, we use 15 values to the left and to the right of each position (using more points to calculate the median is reducing the confidence of the results, since some galaxies do not have enough pixels detected in some bins) and we show the results for each relation over 20 positions across the horizontal axis. The shadowed region surrounding each curve is the median absolute deviation at each position. The vertical axis of each panel spans 4 dex to easily compare the galaxy-to-galaxy variations between resolved SRs.}
    \label{fig:scaling_relations_running_medians_all_sample}
\end{figure*}

Following the strategy from \cite{Ellison2021a} and \cite{Abdurrouf2022b}, in \cref{fig:scaling_relations_running_medians_all_sample} we calculate the running median to the detected values of individual galaxies (coloured lines), as well as for the whole sample (black lines). The rMGMS (mid panel) and rSK (left panel) are displaying lower values of median absolute deviation (MAD), with mean values of $0.14$ and $0.18$\,dex, respectively, and $0.26$\,dex for the rSFMS (right panel). We also apply an ODR linear fit to the full-sample running medians (dashed black lines; parameters are available in \cref{tab:ODR_fits_running_medians}), and we observe that the trends of the running medians are actually not perfectly straight, but they are instead curves with increasing slope along the x-axis (especially the rSK and rSFMS).

\begin{figure}[ht]
    \centering
    \includegraphics[width=0.8\hsize]{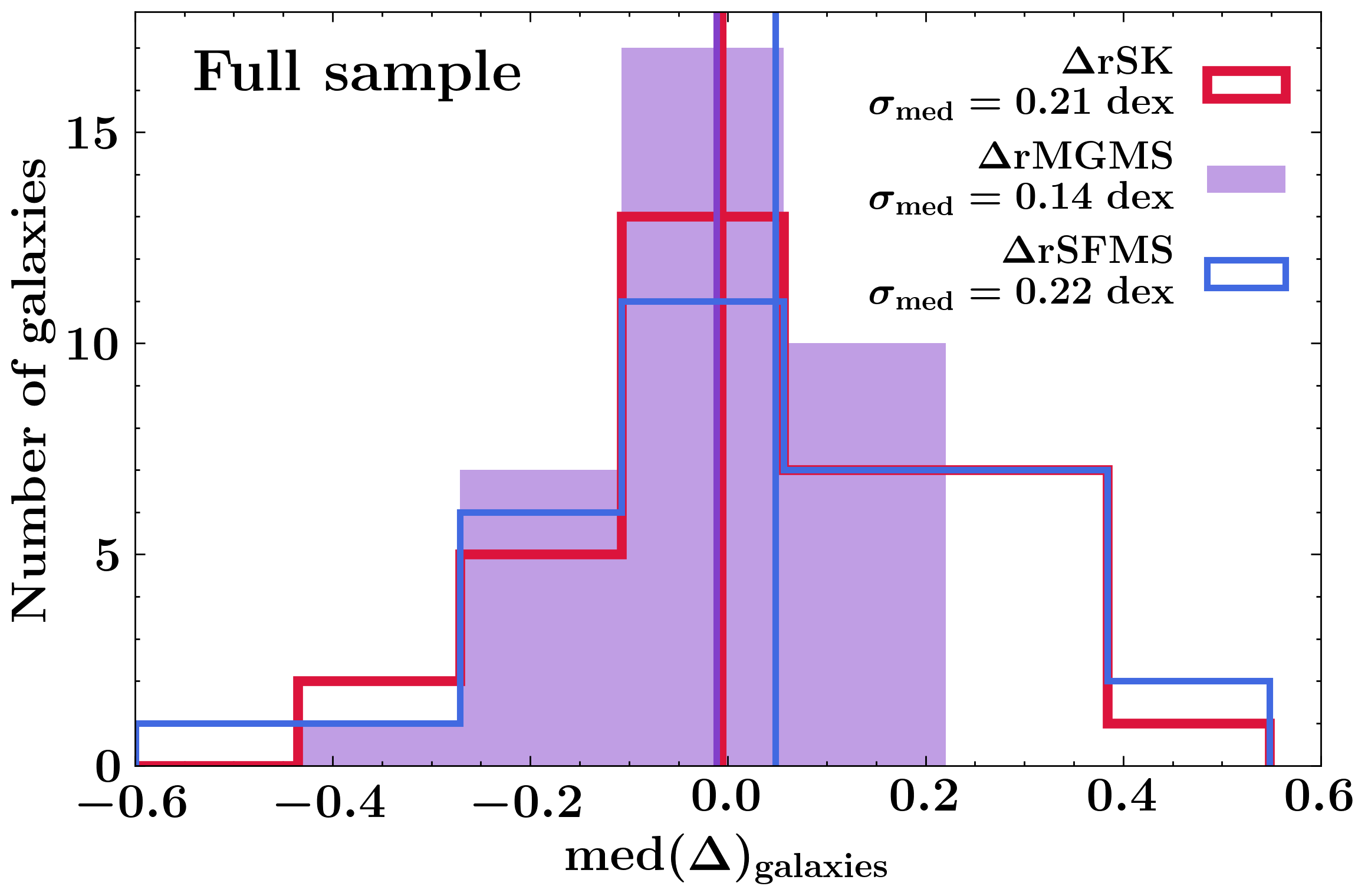}
    \includegraphics[width=0.8\hsize]{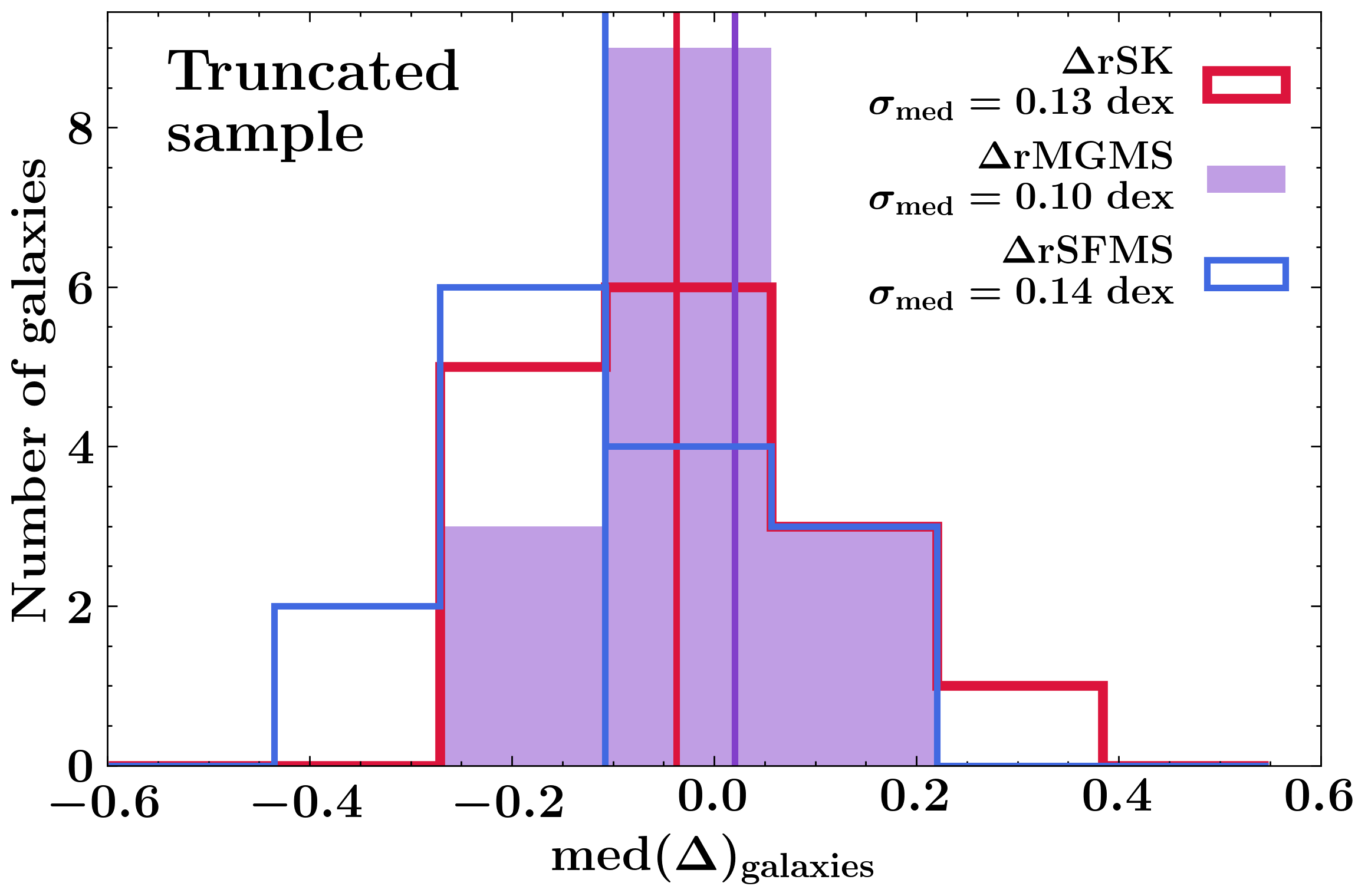}
    \caption{Median deviations of individual galaxies in the resolved SRs with respect to the running median of the full-sample distribution of pixels. In the top panel we show the distributions of medians of $\rm \Delta rSK$, $\rm \Delta rSFMS$ and $\rm \Delta rSFMS$ for the complete sample of 35 detected galaxies. In the bottom panel we show the same distributions for the truncated sample of 15 galaxies (see text for explanation). In each panel, we show the median value of the distributions with the red ($\rm \Delta rSK$), purple ($\rm \Delta rMGMS$), and blue ($\rm \Delta rSFMS$) vertical lines. In the upper-right legend of each panel, we also include the standard deviation of each distribution.}
    \label{fig:delta_rSRs_med_gals_hists}
\end{figure}

Once we define these curves, for each galaxy with the curve $y(x)$, we calculate the deviation from the full-sample curve $Y(x)$ as $\Delta(x) = y(x) - Y(x)$, where $x$ is the mass surface density ($\Sigma_\mathrm{H_2}$ for the rSK and $\Sigma_\star$ for the rMGMS and the rSFMS) bin at which the deviation is estimated. In the top panel of \cref{fig:delta_rSRs_med_gals_hists} we represent the distributions of the medians of $\rm \Delta rSK$, $\rm \Delta rSFMS$ and $\rm \Delta rSFMS$ for all the detected galaxies in the sample.

For each relation we show the median value of $\Delta(x)$, as well as the standard deviation of each distribution, $\sigma_\mathrm{med}$. The three relations have median values of $-0.01$\,dex for the rSK and the rMGMS, and $0.05$\,dex for the rSFMS. We obtain a standard deviation of $0.21$\,dex for the rSK, $0.14$\,dex for the rMGMS and $0.22$\,dex for the rSFMS. These results show that the rMGMS is the relation with less deviations with respect to the full-sample distribution, and agrees with the findings of \cite{Ellison2021a}. As previously stated, the rMGMS is quantitatively more consistent and homogeneous for the galaxies of the ALMA CO-CAVITY sample, whereas the rSFMS shows the largest variation between galaxies.

We showed that the galaxy-to-galaxy variations play an important role in the deviations of the relation, and that the rMGMS is less affected by them. To confirm this observation, we re-analyse the results obtained in the previous subsection using exclusively galaxies whose distributions follow that of the full-sample (from the first category presented in \cref{sec:groups_rSRs}).

For this new subsample, we select the 15 galaxies that follow the full-sample trend for the three resolved SRs (hereafter, the `truncated sample'). In \cref{fig:rSK_all_galaxies,fig:rMGMS_all_galaxies,fig:rSFMS_all_galaxies} we show these 15 galaxies' names with a green background colour, and in \cref{fig:resolved_scaling_relations_ODR_fits_distros_all_vs_truncated_sample} we show (top panels) the rSK, rMGMS and rSFMS obtained for this truncated subsample of galaxies (black contours) and compare them with the original distributions (coloured contours). As expected, the distributions have similar contours and fits, although with a reduced number of pixels (2\,669 pixels instead of the original 7\,937 detected pixels). The individual distributions of surface densities visible in the bottom panels also display similar shapes.

The 1$\sigma$ scatters of the ODR fits decrease from 0.21 to 0.17\,dex ($-0.04$\,dex) for the rSK and from and from 0.24 to 0.19\,dex ($-0.05$\,dex) for the rSFMS, while it remains unchanged for the rMGMS at 0.16\,dex. The pixels that were removed represented deviations in the rSK and rSFMS that increased the scatter of the data when they were fitted. On the other side, the rMGMS scatter does not change, suggesting that it is less susceptible to the variations created by these galaxies in the first place.

In the bottom panel of \cref{fig:delta_rSRs_med_gals_hists} we reproduce the analysis made in the top panel of the same figure using exclusively the truncated sample of 15 galaxies. As expected, removing the pixels of galaxies whose resolved SRs deviate from the full-sample distribution results in less standard deviation in the distribution of the median deviations ($\rm med(\Delta)_{galaxies}$). In the case of the rSK, the value of $\sigma_\mathrm{med}$ decreases from 0.21 to 0.13\,dex ($-0.08$\,dex), in the case of the rSFMS, from 0.22 to 0.14\,dex ($-0.08$\,dex) and in the case of the rMGMS, from 0.14 to 0.10\,dex ($-0.04$\,dex).

\subsection{Correlations between deviations}\label{sec:deviations_rSRs_correlated}
\begin{figure*}
    \centering
    \includegraphics[width=\textwidth]{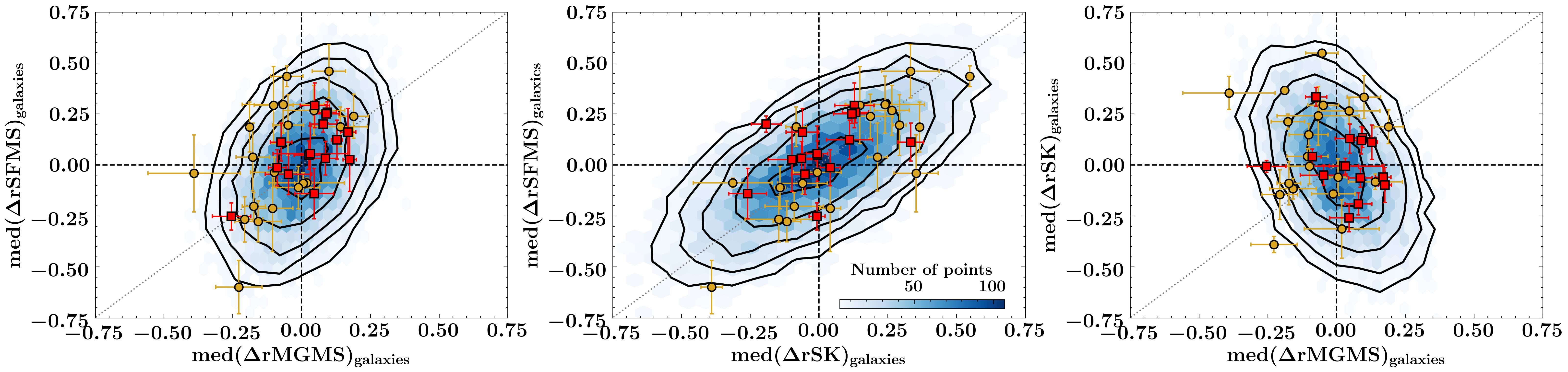}
    \caption{Galaxy-by-galaxy offsets from the full-sample trends of the resolved SRs compared between each relation. In each panel the data points are the the median offsets, with the $\mathrm{rSFMS-rMGMS}$ diagram in the left panel, the $\mathrm{rSFMS-rSK}$ diagram in the mid panel, and the $\mathrm{rSK-rMGMS}$ diagram in the right panel. The error bars are the median absolute deviation of each value. In red we highlight the 15 galaxies of the truncated sample (see text for explanation), and in yellow the other 20. The blue distribution in the background represents the deviations for the mock sample of 10\,000 data points, with black contours depicting 5, 10, 20, 40, and 70\% of inclusion. In each panel the horizontal and vertical black-dashed lines representing the positions of no offset at 0.0\,dex, and with the dotted grey line the 1-to-1 relation.}
    \label{fig:delta_rSRs_med_gals_corr_plot_all_truncated_mock_sample}
\end{figure*}

We aim to understand the underlying connection between the three resolved SRs. In \cref{fig:delta_rSRs_med_gals_corr_plot_all_truncated_mock_sample} we compare the values of $\rm med(\Delta)$ for each SR for the detected galaxy sample. We highlight in red the truncated sample. \cref{tab:Pe_corr_coeffs} summarises the correlation coefficients obtained in each case.

\begin{table}
    \centering
    \caption{Pearson's correlation coefficients between the median offsets of relations from the sample of 35 detected galaxies (full sample), the truncated sample of 15 galaxies, and the mock sample of 10\,000 values (defined in \cref{sec:deviations_rSRs_correlated}).}
    \resizebox{\hsize}{!}{
    \begin{tabular}{cccc}
        \hline\hline
                  & $\rm \Delta rMGMS - \Delta rSFMS$ & $\rm \Delta rSK - \Delta rSFMS$ & $\rm \Delta rSK - \Delta rMGMS$ \\ \hline
        Full sample & \begin{tabular}[c]{@{}c@{}}0.52\\ ($1.37\times10^{-3}$)\end{tabular} & \begin{tabular}[c]{@{}c@{}}0.70\\ ($3.10\times10^{-6}$)\end{tabular} & \begin{tabular}[c]{@{}c@{}}$-0.10$\\ (0.58)\end{tabular} \\ \hline
        
        Truncated sample & \begin{tabular}[c]{@{}c@{}}0.54\\ ($3.80\times10^{-2}$)\end{tabular} & \begin{tabular}[c]{@{}c@{}}0.39\\ ($0.15$)\end{tabular} & \begin{tabular}[c]{@{}c@{}}$-0.35$\\ (0.20)\end{tabular} \\ \hline
        
        Mock sample & 0.41 & 0.65 & $-0.42$ \\ \hline\hline
        
    \end{tabular}}
    \tablefoot{
    The correlation coefficient between the median offsets of the relations is shown in each cell, with the $\mathrm{p-value}$ in parentheses below. In the case of the mock sample, the $\mathrm{p-value}$ is null.
    }
    \label{tab:Pe_corr_coeffs}
\end{table}

We observe that the median deviations are correlated. The highest correlation is obtained between the median offsets of the rSK with those of the rSFMS (mid panel): this means that variations in one of these relation strongly affect the other. There is a weaker correlation between the deviations of the rMGMS and those of the rSFMS (left panel), showing that variations in the content of gas and $M_\star$, that affect the rMGMS, will also affect the $\Sigma_\mathrm{SFR}$ used in the rSFMS, but less than variations in the rSK. On the other side, a weak anti-correlation is visible between the deviations of the rSK and those of the rMGMS (right panel). These results are confirmed when looking at the deviations from the truncated sample, as the anti-correlation in the right panel is stronger with this sub-sample.

A natural question that arises is whether these correlations are physically driven. They could be the result of the covariances between each quantity due to their respective errors, or they could be associated with selection effects, or one SR could be linearly dependent on the other two. This was actually observed in the case of EDGE-CALIFA~\citep[e.g.][]{Sanchez2021,GaraySolis2024}, where correlations between the offsets of the quantities with respect to the SRs were proven to be artificial and not physical. To determine if the same is happening to the ALMA CO-CAVITY sample, we apply a similar methodology as \citep{Sanchez2021}.

We create mock data for 10\,000 values, sampling the $\Sigma_\star$ axis following a Gaussian distribution centred on the mean value of $\log(\Sigma_\star/\mathrm{[M_\odot\ kpc^{-2}]}) = 7.94$ and its width as the standard deviation (0.40\,dex). We then use the parameters from \cref{tab:rSRs_all_fits_full_sample_surveys} to create values of $\Sigma_\mathrm{H_2}$ (using the rMGMS) and $\Sigma_\mathrm{SFR}$ (using the rSFMS). We finally add a random noise following a Gaussian distribution with a width of 0.1\,dex for $\Sigma_\star$ and $\Sigma_\mathrm{H_2}$, and 0.2\,dex for $\Sigma_\mathrm{SFR}$, as the median uncertainty of the observed $\Sigma_\mathrm{SFR}$ is twice that of $\Sigma_\mathrm{H_2}$ and $\Sigma_\star$ (0.09, 0.05 and 0.05\,dex, respectively).

To estimate the offsets from the mock data, we simply calculate the vertical distance from each value in the rSK, rMGMS and rSFMS to its respective linear fit. A difference is that we did not group by individual galaxies as naturally in our data, but the same result would emerge. In \cref{fig:delta_rSRs_med_gals_corr_plot_all_truncated_mock_sample} we show in the background the distributions of offsets from the mock sample. We see that they coincide with the observed data points, and the Pearson's correlation coefficients from \cref{tab:Pe_corr_coeffs} confirm the trends we previously commented.

\section{Discussion}\label{sec:discussion}
The main focus of this work is to study whether the resolved SRs depend on the LSE. We compare the resolved SRs of our sample of VGs with other samples, composed of galaxies located in denser environments. The results of \cref{sec:full_sample_rSRs} show that the linear ODR fits of the relations are similar in voids to that of galaxies from denser LSEs, in particular from ALMaQUEST XI full sample (\cref{fig:triple_general_plot_ODR}). This shows that the SRs are fundamental from the point of view of the LSE, in the sense that, regardless of the environment, similar relations are found. This was also shown in \citetalias{Espada2026}, although globally.

Nonetheless, contrary to our comparison surveys, we find the rMGMS (and not the rSK) to be the tightest of the three relations (\cref{fig:triple_general_plot_ODR}). This could be well explained by the fact that both ALMaQUEST and EDGE-CALIFA samples encompass galaxies across all types of LSEs, whereas we only cover VGs. A potential explanation comes from the intrinsic differences between the galaxies in our sample and those studied in previous surveys: the predominant isolated local environment in voids. Most galaxies in the ALMA CO-CAVITY sample (83\%) are singlets, and 65\% are isolated (34 and 27 galaxies respectively; \citetalias{Espada2026}), whereas in other samples such as ALMaQUEST and EDGE-CALIFA, this number decreases to $20-25$\% of isolated galaxies. These percentages are in agreement with what is found for galaxies in the Local Universe, where 78\% of VGs are singlets, whereas this value decreases to 22\% for cluster galaxies and 60\% for galaxies not belonging to either of these two LSEs (galaxies not in voids nor clusters, NCNV, \citealt{TorresRios2026,ArgudoFernandez2026})\footnote{These values are obtained considering only galaxies with a mass $\log_{10}(M_\star/\mathrm{M_\odot})>9.5$, for the sake of completeness, which means that these percentages are upper limits since some satellites might have not been considered.}.

Interactions between galaxies rich in gas are often translated into increased $\Sigma_\mathrm{H_2}$ and $\Sigma_\mathrm{SFR}$, enlarging the dynamic range of the rSK \citep[e.g.][]{Teyssier2010}. This would explain why surveys in denser LSEs, such as in ALMaQUEST XI, find a stronger correlation of the rSK. Due to the small environmental density and large degree of isolation of the galaxies in the cosmic voids, the dynamic range in the rSK is smaller. On the other hand, the correlation between $\Sigma_\star$ and $\Sigma_\mathrm{H_2}$ is maintained, whereas in other local and LSEs the scatter of the rMGMS increases.

As mentioned in \cref{sec:introduction}, the rSK is a fundamental relation induced by the formation of stars in giant molecular clouds. The relation between $\Sigma_\mathrm{H_2}$ and $\Sigma_\star$ in the rMGMS is not as trivial. Works like the one carried out by \cite{Baker2022} (and previously \citealt{Wong2013,Lin2019}) suggest that either the distribution of molecular gas aligns with the gravitational potential of the stellar distribution, or that both are following the gravitational potential of the dark matter, if it represents a meaningful fraction of the dynamical mass. Following this, the rMGMS in our sample, dominated by isolated galaxies, is tighter, indicating a stronger correlation between stellar and molecular gas content. This is possibly due to the lower fraction of events in our sample, such as mergers or starbursts, that would disrupt the gravitational potential of galaxies.

We saw in \cref{sec:individual_rSRs} that a large fraction of the dispersion in the fits to the relations results from variations between galaxies, similarly to what previous studies have shown \citep[e.g.][]{Lin2019,Ellison2021a}, and also from local variations within galaxies associated with different substructures (\cref{sec:rSK_individual}), for example, strong bars (such as \texttt{58154}), or spiral arms (\texttt{59266}). In \cref{sec:measuring_offsets_rSRs}, our results confirm that the rMGMS is less susceptible to galaxy-to-galaxy variations, as the distribution of the offsets with respect to the global trend changed less once we remove the galaxies with large variations. We also observe that, by removing these galaxies, the standard deviation of the median offsets in the rSK is much closer to that of the rMGMS: this suggests that the additional offsets in the rSK with respect to the rMGMS can be attributed primarily to galaxy-to-galaxy variations, and that once these variations are ignored, the rSK can be as homogeneous (or at least much more) as the rMGMS between VGs. A future paper of the ALMA CO-CAVITY series will try to disentangle the reasons behind the galaxy-to-galaxy variations in VGs.
 
Additionally, the visual classification of the resolved SRs (\cref{sec:app_categories_galaxies_rSRs}) demonstrated that the rSK and rSFMS tend to present similar behaviours in each group, whereas the rMGMS is behaving differently. We argue that $\Sigma_\star$ and $\Sigma_\mathrm{H_2}$ are better correlated than the variables involved in the other two relations because changes in SFR may occur in magnitude and on time-scales too short to be detected in either of the other two quantities. As mentioned in \cref{sec:full_sample_rSRs}, $\Sigma_\mathrm{H_2}$ is calculated from CO(1--0), which traces the bulk of molecular gas, but it is not as related to the SF as the denser gas that transforms into stars. $\Sigma_\star$ is directly tracing the old stellar populations of the galaxies, which means that, while the rSK and rSFMS are able to detect rapid changes in the SF (from interactions or starburst, of the order of hundreds of Myr, e.g. \citealt{DiMatteo2008}), the rMGMS traces changes on a longer time-scale and is less susceptible to the changes in SF. Furthermore, the scatter of the resolved SRs is mainly caused by galaxy-to-galaxy variations, which supports the idea that the changes in SF are mainly occurring at the galactic scale.

Finally, regarding the median offsets of the relations, we observe correlations between the rMGMS and rSK with respect to the rSFMS, and an anti-correlation between the rMGMS and rSK. However, the analysis with the mock sample confirms that the correlations naturally emerge from the parametrisation of the resolved SRs and the dispersion (in this case parametrised as a random Gaussian noise), meaning that they cannot be attributed entirely to the physical relations between the rSK, rMGMS, and rSFMS. They are a consequence of the projection of the 3-D primary relations onto the planes defined by the paired combinations of the $\Sigma_\star$, $\Sigma_\mathrm{H_2}$ and $\Sigma_\mathrm{SFR}$ axes, as studied by \cite{Lin2019,Sanchez2021}. Our results coincide with that of EDGE-CALIFA~\citep{Sanchez2021,GaraySolis2024} and no conclusion can be drawn towards the existence of physical secondary relations between the residuals of SRs in VGs. Based on the correlation coefficients, the values for the full sample are higher than those of the mock sample, and we cannot exclude a possible physical correlation between the deviations. A strong relation $\Sigma_\mathrm{SFR}\propto\Sigma_{\star}^\alpha\Sigma_\mathrm{H_2}^\beta$ has previously been observed in other studies \citep[e.g.][]{Sanchez2021}, and it can be related to physical quantities such as the dynamical equilibrium pressure \citep[e.g.][]{BarreraBallesteros2021,Ellison2024}. Future work will examine this relation in the 3-D parameter space of the ALMA CO-CAVITY VGs to determine whether additional differences can be identified between the SRs of VGs and in other LSEs.

\section{Summary and conclusions}\label{sec:conclusions}
In this work we present the rSK, rMGMS and rSFMS for 41 VGs, at a resolution of 2\farcs5, corresponding to linear scales of 0.8--2.1\,kpc. We used interferometric data from the ALMA observatory to study the distribution of molecular gas in these galaxies. We also used IFU data from the CAVITY and MaNGA surveys to obtain ionised gas and stellar population distributions, and we combine all of them to study the resolved SRs in VGs. Below we summarise the results of this study:

   \begin{itemize}[leftmargin=0pt, itemsep=0pt, topsep=0pt, label={}]
      \item $-$ The results of the ODR linear fits of the rSK, rMGMS, and rSFMS in VGs are similar than in other surveys, showing that these relations are fundamental across LSEs.
      
      \item $-$ We find for the full-sample of detected pixels (\cref{fig:triple_general_plot_ODR}), a super-linear parametrisation for the rSK (with a slope of $1.282\pm0.011$, corresponding to $\tau_\mathrm{dep}\sim1-2$\,Gyr) and rSFMS ($1.249\pm0.013$), and a linear fit for the rMGMS ($0.955\pm0.007$). These parameters agree with the fits from other samples including galaxies in denser environments (e.g. ALMaQUEST XI).
      
      \item $-$ The smallest scatter is found for the rMGMS, with 0.16\,dex, then the rSK with 0.21\,dex, and finally the rSFMS with 0.24\,dex. This differs from previous studies such as ALMaQUEST XI and EDGE-CALIFA, that encompass galaxies across all LSEs, where the rSK presents the smallest scatter. We argue that this difference may arise from the isolation of the VGs in our sample. In contrast, the galaxies in other samples may have experienced more interactions and changes in SF (such as starbursts) that have altered the distributions of surface densities.
      
      \item $-$ When looking at the relations of individual galaxies (\cref{fig:rSK_all_galaxies,fig:rMGMS_all_galaxies,fig:rSFMS_all_galaxies}), we observe a large galaxy-to-galaxy variability, similarly to what is observed in previous studies including other samples. This is especially the case for the rSK and the rSFMS.
      
      \item $-$ Despite the galaxy-to-galaxy variability, some patterns are visible (we show several examples in \cref{fig:resolved_scaling_relations_galaxy_categories_representative}), such as relations saturating, with several trends depending on the location of the pixel in the disc, or below or above the full-sample distribution.
      
      \item $-$ When we focus on the galaxies whose relations are above or below the full-sample distribution, we find several groups. In all of them, the rMGMS behaves differently than the other two relations, indicating that what is causing the variations in the resolved SRs affects differently the rMGMS.
      
      \item $-$ Galaxies without detected molecular gas also present similarities, such as their $M_\star$ around $\rm 10^{9.2-9.4}\ M_\odot$, and their blue and late-type morphology, which raises the question of why no molecular gas is detected. For three of them (half of the sample), their tight rSFMS shows distinctive shapes, possibly due to a suppression of their SF, but the reasons behind this phenomenon will require further investigation.
      
      \item $-$ Applying a preliminary analysis on the deviations suffered by the resolved SRs (\cref{fig:scaling_relations_running_medians_all_sample}), we confirm that the rMGMS is less affected by the galaxy-to-galaxy variations (\cref{fig:delta_rSRs_med_gals_hists}). When removing those galaxies that deviate from the full-sample distribution, the rMGMS remains the tightest SR and the one that changed the least. We hypothesise that this is due to the fact that the rMGMS is intrinsically different than the rSK and the rSFMS, as it does not trace the SF (or changes in the SF) directly. We also argue that the changes in SF are mainly occurring at galactic scales since the deviations to the resolved SRs are caused by the galaxy-to-galaxy variations.
      
      \item $-$ The deviations found between the rSK and the rSFMS, and between the rMGMS and the rSFMS, are correlated (\cref{fig:delta_rSRs_med_gals_corr_plot_all_truncated_mock_sample}). However, this result cannot be associated to a physical relation between the resolved SRs, as it can be reproduced by introducing Gaussian noise in the relations. It results from the projection of the three-dimensional relation between $\Sigma_\star$, $\Sigma_\mathrm{H_2}$ and $\Sigma_\mathrm{SFR}$. This coincides with results in previous surveys \citep[e.g. EDGE-CALIFA][]{Sanchez2021}.
   \end{itemize}

The study of the SRs is an ongoing area of research. In this study, we targeted VGs and our findings suggest that while the general SRs remain valid, the environment may play a role on the variations found in SRs and, ultimately, on the evolution of the galaxies. We analysed the relations between $\Sigma_\mathrm{H_2}$, $\Sigma_\star$ and $\Sigma_\mathrm{SFR}$, but additional studies also found a correlation, in some cases tighter than with $\Sigma_\mathrm{H_2}$, with the dense-gas phase traced by HCN \citep[e.g.][]{Wu2010,JimenezDonaire2019}, the dynamical equilibrium pressure \citep{BarreraBallesteros2021,Ellison2024,Corbelli2025}, or the dust \citep[e.g.][]{Abdurrouf2022b,Salvestrini2025}. This encourages us to progress towards a better understanding of the resolved SRs, and ALMA CO-CAVITY provides an ideal platform for studying these relations in VGs to disentangle the role of the local and large-scale environments. In future work, we will further exploit the ALMA CO-CAVITY dataset to study these other SRs. Additional analyses, including principal component analysis and random forest regression will provide key insights into the deviations to the resolved SRs, how these deviations relate to the evolutionary paths of individual galaxies, and ultimately how the LSE influences them.

\section{Data availability}\label{sec:data_availability}
The atlas of resolved SRs of the 41 VGs is available as supplementary material. A pipeline of the estimations of the surface density values used in this work ($\Sigma_\star$, $\Sigma_\mathrm{H_2}$, and $\Sigma_\mathrm{SFR}$) is available at \url{https://gitlab.com/astrogal/alma-co-cavity}.

\begin{acknowledgements}
    We thank the anonymous referee for carefully reviewing our manuscript. We truly appreciate the assessment of our work and the constructive comments provided, which have helped us to improve the paper.
    
    This paper is partially based on observations collected at the Centro Astronómico Hispano en Andalucía (CAHA) at Calar Alto, operated jointly by Junta de Andalucía and Consejo Superior de Investigaciones Científicas (IAA-CSIC), under the CAVITY legacy project. The CAVITY project acknowledges financial support by the research projects AYA2017-84897-P, PID2020-113689GB-I00, PID2020-114414GB-I00, and PID2023-149578NB-I00 funded by the Spanish Ministry of Science and Innovation (MCIN/AEI/10.13039/501100011033) and by FEDER/UE; the project A-FQM-510-UGR20 funded by FEDER/Junta de Andalucía-Consejería de Transformación Económica, Industria, Conocimiento y Universidades/Proyecto; by the grants P20-00334 and FQM108, funded by Junta de Andalucía; and by Consejería de Universidad, Investigación e Innovación (Junta de Andalucía) and Gobierno de España and European Union NextGenerationEU through grant AST22\_4.4.

    This paper makes use of the following ALMA data: ADS/JAO.ALMA\#2022.1.00482.S. ALMA is a partnership of ESO (representing its member states), NSF (USA) and NINS (Japan), together with NRC (Canada), MOST and ASIAA (Taiwan), and KASI (Republic of Korea), in cooperation with the Republic of Chile. The Joint ALMA Observatory is operated by ESO, AUI/NRAO and NAOJ.
    
    This project used data obtained with the Dark Energy Camera (DECam), which was constructed by the Dark Energy Survey (DES) collaboration.

    The project leading to this publication has received support from ORP, that is funded by the European Union's Horizon 2020 research and innovation programme under grant agreement No 101004719.

    This project makes use of the MaNGA-Pipe3D dataproducts. We thank the IA-UNAM MaNGA team for creating this catalogue, and the Conacyt Project CB-285080 for supporting them.

    This work made use of the CARTA (Cube Analysis and Rendering Tool for Astronomy) software (DOI \url{https://zenodo.org/records/15172686} – \url{https://cartavis.org}).

    This work made use of the following software packages: \texttt{astropy} \citep{astropy:2022}, \texttt{Jupyter} \citep{2007CSE.....9c..21P,kluyver2016jupyter}, \texttt{matplotlib} \citep{Hunter:2007}, \texttt{numpy} \citep{numpy}, \texttt{pandas} \citep{mckinney-proc-scipy-2010,pandas:10957263}, \texttt{python} \citep{python}, \texttt{scipy} \citep{2020SciPy-NMeth,scipy_11255513}, \texttt{scikit-image} \citep{scikit-image}, \texttt{seaborn} \citep{Waskom2021}, and \texttt{TOPCAT} \citep{2005ASPC..347...29T}. This research made use of the dust extinction models provided by \texttt{dust\_extinction} \citep{Gordon+2024:2024JOSS....9.7023G}. This research made use of Photutils, an Astropy package for detection and photometry of astronomical sources \citep{Photutils_13989456}.

    This research has made use of the NASA/IPAC Extragalactic Database, which is funded by the National Aeronautics and Space Administration and operated by the California Institute of Technology. Funding for SDSS-III has been provided by the Alfred P. Sloan Foundation, the Participating Institutions, the National Science Foundation, and the U.S. Department of Energy Office of Science. The SDSS-III Web site is \url{http://www.sdss3.org/}. The SDSS-IV site is \url{http://www.sdss.org}. This project used data obtained with the Dark Energy Camera (DECam), which was constructed by the Dark Energy Survey (DES) collaboration.

    We acknowledges financial support from PID2025-175276NB-I00 funded by MCIU/AEI/10.13039/501100011033.

    SBD acknowledges financial support from the grant AST22.4.4, funded by Consejería de Universidad, Investigación e Innovación and Gobierno de España and Unión Europea –- NextGenerationEU, project PID--2021--122544NB--C43, project PID2023-170178NB-I00, and the Institut de Radioastronomie Millimétrique (IRAM).
    
    DE acknowledges support from a Beatriz Galindo senior fellowship (BG20/00224) from the Spanish Ministry of Science and Innovation.

    SDP acknowledges financial support by MINECO under grants PID2023-149578NB-100 and PID2022-136598NB-C32.
    
    YGK acknowledges financial support from PREP2023-001684 funded by MCIU/AEI/10.13039/501100011033 and the FSE+.
    
    GTR acknowledges financial support from the research project PRE2021-098736, funded by MCIN/AEI/10.13039/501100011033 and the FSE+.

    SFS acknowledges the support by CBF-2025-I-236 project granted by the Secretaría de Ciencia, Humanidades, Tecnología e Innovación (SECIHTI) of the Mexican Federal Government, and the PID2022-136598NB-C31 (ESTALLIDOS) grant by the Spanish Ministery of Science and Innovation (MCINN).
    
    AC and RGB acknowledge financial support from the Severo Ochoa grant CEX2021-001131-S funded by MCIN/AEI/10.13039/501100011033, and the grant PID2022-141755NB-I00.

    MAF and PVB acknowledge support from the Emergia program (EMERGIA20\_38888) from Consejería de Universidad, Investigación e Innovación de la Junta de Andalucía.

    TRL acknowledges support from Ram\'on y Cajal fellowship (RYC2023-043063-I, financed by MCIU/AEI/10.13039/501100011033 and by the FSE+).

    LSM, AZ and EF acknowledge support from PID2023-150178NB-I00, financed by MCIU/AEI/10.13039/501100011033, and by FEDER, UE.
\end{acknowledgements}

\bibliographystyle{aa.bst}
\bibliography{reference}

\begin{appendix}
\nolinenumbers
\section{Effect of local $\alpha_\mathrm{CO}$ on the scaling relations}\label{sec:app_variable_local_alphaCO}
\begin{figure}[h!]
    \centering
    \includegraphics[width=\hsize]{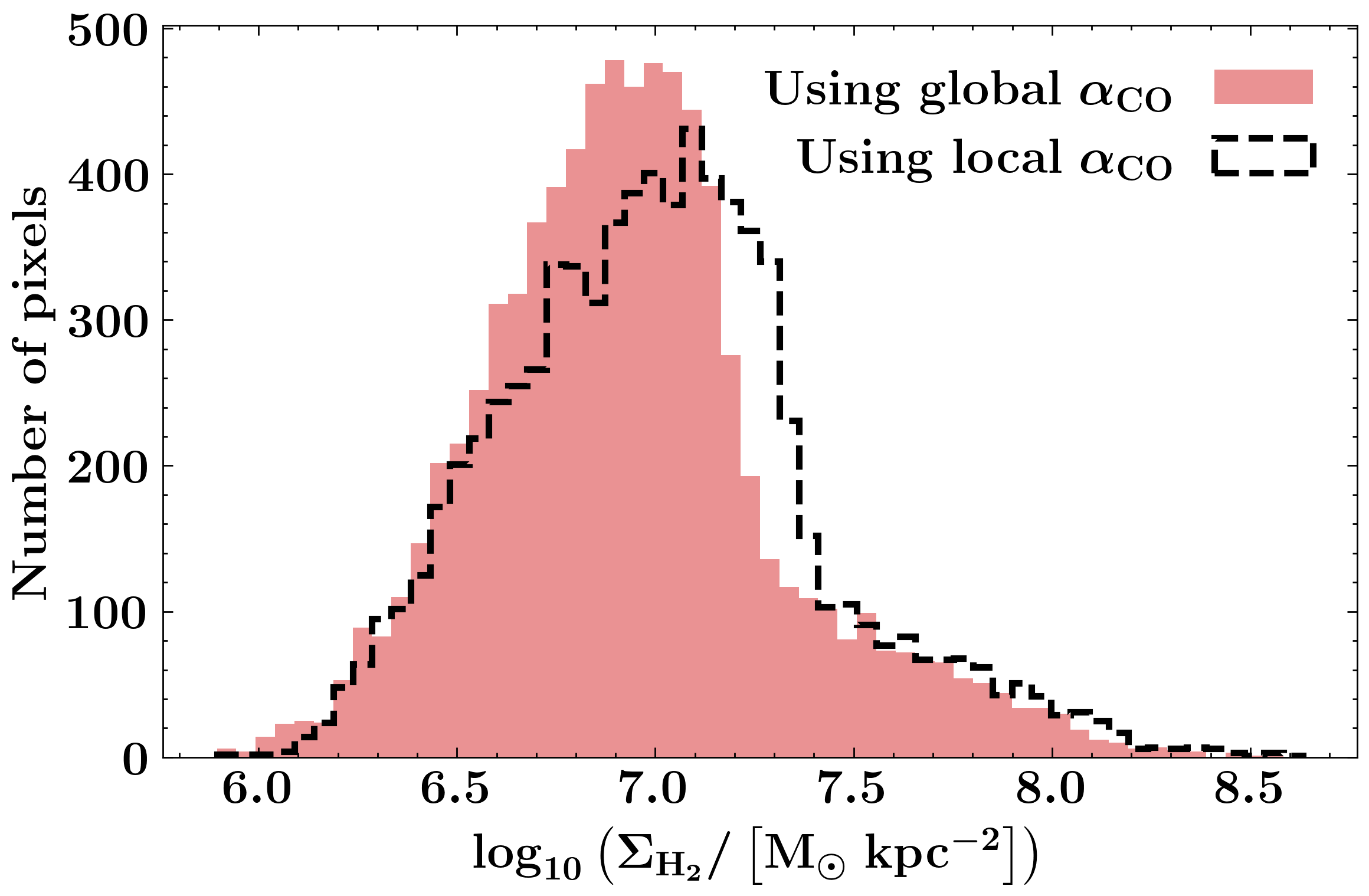}
    \caption{Surface densities distributions of $\Sigma_\mathrm{H_2}$, estimated using global and local $\alpha_\mathrm{CO}$ values.}
    \label{fig:resolved_surface_densities_histogram_SigmaH2_local_alphaCO}
\end{figure}

During the analysis of the data we used a global $\alpha_\mathrm{CO}$ conversion factor to calculate $\Sigma_\mathrm{H_2}$. In this appendix we discuss the results obtained using a local $\alpha_\mathrm{CO}$ conversion factor. In \cref{fig:resolved_surface_densities_histogram_SigmaH2_local_alphaCO} we show the distributions of $\Sigma_\mathrm{H_2}$ calculated using the global value of $\alpha_\mathrm{CO}$ applied throughout the paper, and the local values of $\alpha_\mathrm{CO}$. The values of $\Sigma_\mathrm{H_2}$ slightly increased, with a mean $\log_{10}(\Sigma_\mathrm{H_2}/[\mathrm{M_\odot\ kpc^{-2}}])$ value increasing from 6.94 to 7.01, while the standard deviation increased from 0.38\,dex to 0.40\,dex.

\begin{figure*}
    \centering
    \includegraphics[width=\textwidth]{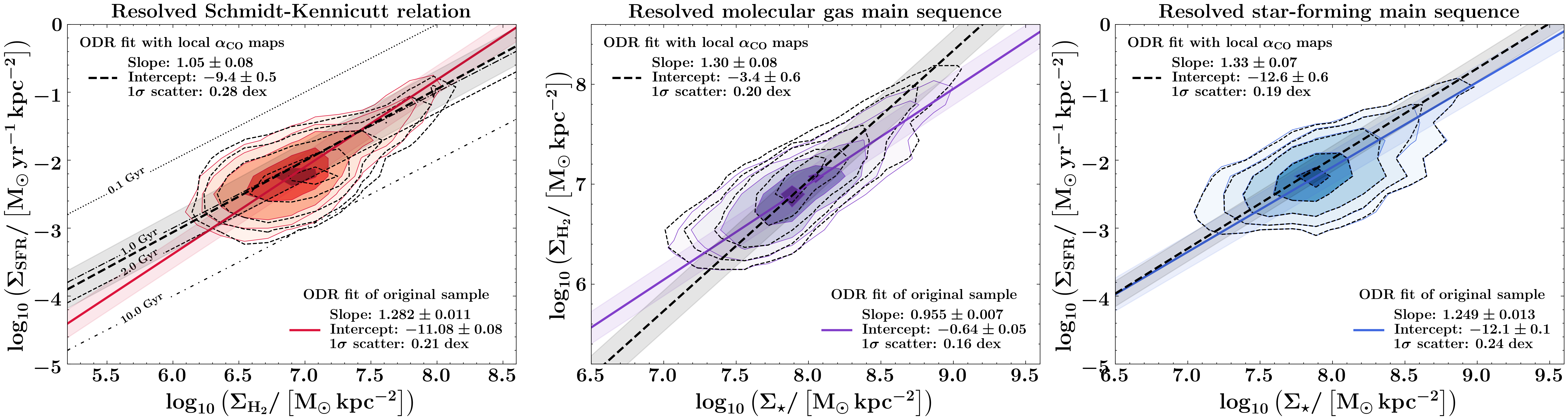}
    \caption{Resolved SRs for the 41 ALMA CO-CAVITY galaxies, using local $\alpha_\mathrm{CO}$-estimated $\Sigma_\mathrm{H_2}$. We show with coloured contours the distribution of the original sample of 7\,937 pixels, and with black-dashed contours the distributions obtained using the local $\alpha_\mathrm{CO}$-estimated $\Sigma_\mathrm{H_2}$. On top of each distribution we show the corresponding ODR fit, whose parameters are visible in the upper-left and lower-right corners of each panel. In the case of the rSFMS (right panel), we note that the fit and contours vary due to the updated number of detected values in the three axes.}
    \label{fig:triple_general_plot_ODR_local_alphaCO}
\end{figure*}

To determine if the use of local $\alpha_\mathrm{CO}$ can affect the results, we will focus on the resolved SRs obtained using the full sample of detected pixels. In \cref{fig:triple_general_plot_ODR_local_alphaCO} we show the three resolved SRs obtained using the original sample of detected pixels (coloured contours) and the new values of $\Sigma_\mathrm{H_2}$ estimated using a local $\alpha_\mathrm{CO}$ (black dashed contours). We also show the ODR fit obtained with each distribution, with the fitted parameters in the lower-right and upper-left corners, respectively.

The results slightly change for the first two relations, while they remain very similar for the rSFMS, as it does not depend on $\Sigma_\mathrm{H_2}$ values. We see that, despite the increment in the values of $\Sigma_\mathrm{H_2}$, the contours align with the original distribution and, most importantly, the fits remain very similar and within the uncertainty areas depicted by the shadowed areas. Finally, the rMGMS remains the tightest relation, followed by the rSK and the rSFMS, confirming that the results we obtained are not affected by the selection of a local $\alpha_\mathrm{CO}$ parameter.

\section{Data homogenisation example}\label{sec:app_data_homogenisation_example}
\begin{figure}
    \centering
    \includegraphics[width=\hsize]{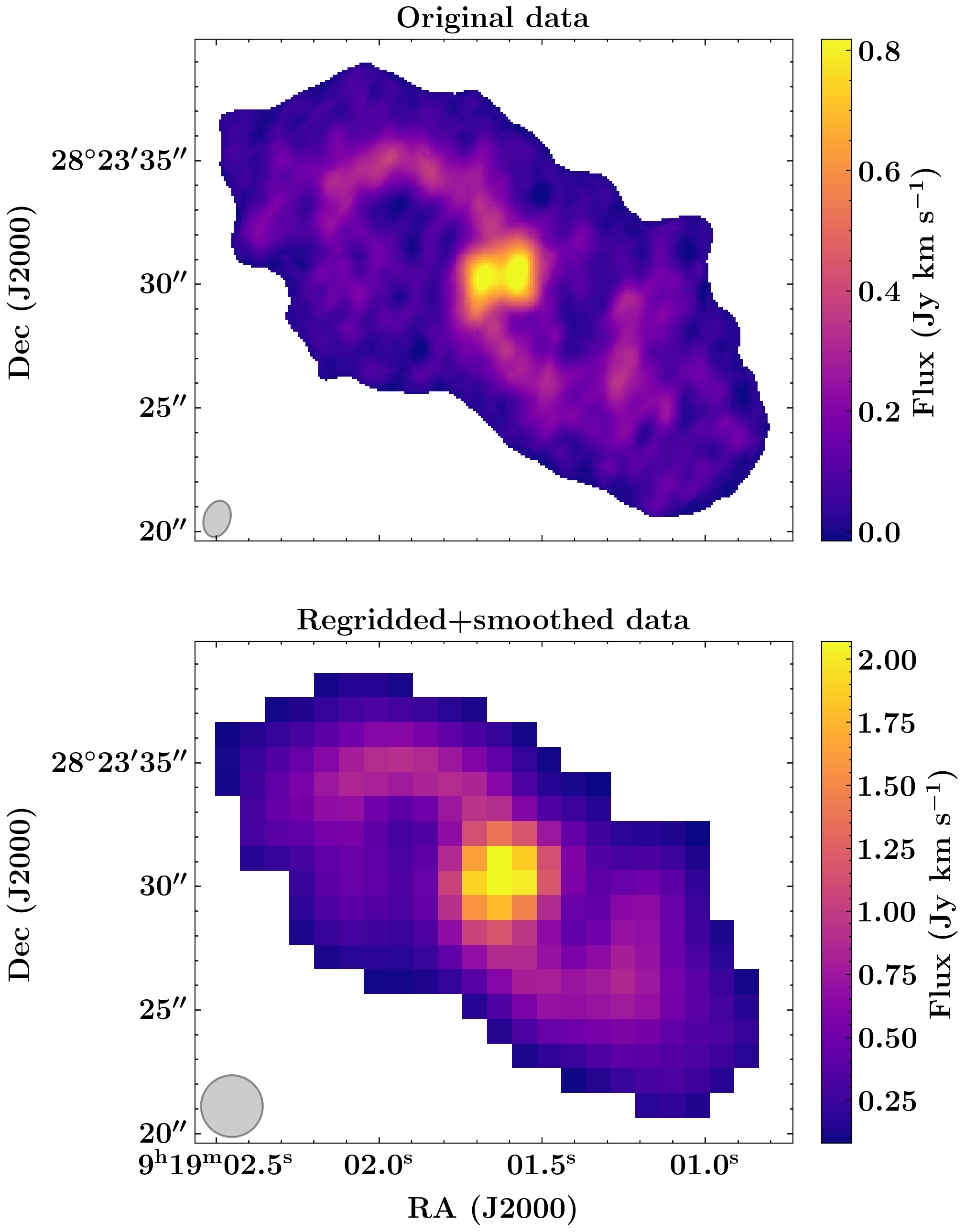}
    \caption{(Top) Original and (bottom) regridded+smoothed CO(1--0) moment-0 map for galaxy \texttt{48125}. We show how the regridding and smoothing of the data are applied to the moment-0 maps, to reconcile the resolution and pixel size of ALMA data with optical data. In this particular case, the data were smoothed to the resolution of 2\farcs5 and the regridding was applied to obtain a pixel size of 1\farcs0, to match PPak data characteristics. We show the restoring beam of the data in the lower-left corner of each panel.}
    \label{fig:smoothing_example}
\end{figure}

By regridding and smoothing the ALMA CO-CAVITY data, we obtain maps with homogenised resolution of $2\farcs5$ and pixel scale of $0\farcs5$ for MaNGA galaxies and $1\farcs0$ for PPak galaxies. We show an example of the results of regridding and smoothing the CO(1--0) moment-0 map of one PPak galaxies in \cref{fig:smoothing_example}.

\section{ODR fits of resolved scaling relations for pixels detected in two of the three axes}\label{sec:app_rSRs_23axes}

The relations shown in \cref{fig:triple_general_plot_ODR} are based on the distribution of pixels detected in the three axes ($\Sigma_\mathrm{H_2}$, $\Sigma_{\star}$, and $\Sigma_\mathrm{SFR}$). However, it is possible to look at the relations taking into account pixels detected in only the two axes of each diagram. For instance, we count 7\,943 pixels detected in both $\Sigma_\mathrm{H_2}$ and $\Sigma_\mathrm{SFR}$ axes ($+0.1\%$), 8\,244 pixels detected in the $\Sigma_\star$ and $\Sigma_\mathrm{H_2}$ axes ($+3.9\%)$, and 19\,668 pixels detected in the $\Sigma_\star$ and $\Sigma_\mathrm{SFR}$ axes ($+147.8\%$).

\begin{figure*}
    \centering
    \includegraphics[width=\textwidth]{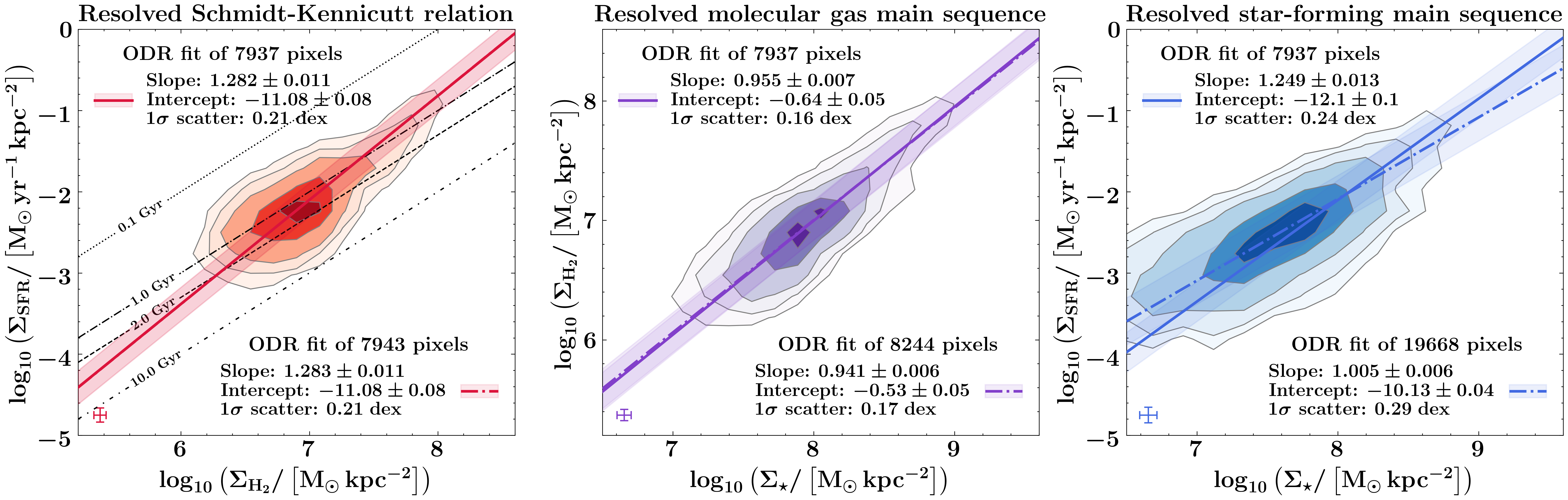}
    \caption{Resolved SRs for the 41 ALMA CO-CAVITY galaxies, similar to \cref{fig:triple_general_plot_ODR}, using pixels detected in two of the three axes. For each relation, we show the pixels detected in its two respective axes ($\Sigma_\mathrm{H_2}$ and $\Sigma_\mathrm{SFR}$ for the rSK, $\Sigma_\star$ and $\Sigma_\mathrm{H_2}$ for the rMGMS, and $\Sigma_\star$ and $\Sigma_\mathrm{SFR}$ for the rSFMS). In the rSK, we count 7\,943 detected pixels, in the rMGMS 8\,244 pixels, and in the rSFMS 19\,668 pixels. In each panel, we show in the lower-left corner the median of the error bars for each distribution. For each relation, we apply an ODR fit which results are shown in the lower-right legend of each panel and are represented by the red (rSK), purple (rMGMS), and blue (rSFMS) dash-dotted lines. For comparison, we also include the fits of the original sample of 7\,937 pixels with the solid lines, and its parameters in the upper-left corner of the panels. In both cases, the vertical scatter is shown as a shadowed region surrounding each fit.}
    \label{fig:triple_general_plot_extended}
\end{figure*}

When including these additional pixels to the diagrams of the relations, we obtain the results shown in \cref{fig:triple_general_plot_extended}. The rSK and rMGMS do not change drastically, with similar fits and scatters, due to the small change in the number of additional pixels included. However, the shape of the rSFMS changes drastically, with $>1.4$ times more pixels than in the original sample. It expanded towards lower values of $\Sigma_\star$, reaching a plateau around $\rm \log_{10}\left(\Sigma_\mathrm{SFR}/\left[M_\odot\ yr^{-1}\ kpc^{-2}\right]\right) \sim -3.5$. This shows that the dynamic range of the $\Sigma_\star$ axis is actually larger than what is shown in \cref{fig:triple_general_plot_ODR}, with values much below $\rm 10^{7.0}\ M_\odot\ kpc^{-2}$. Furthermore, the slope of the fit is $1.005\pm0.006$, getting closer to linearity, but the scatter is only increasing 0.05\,dex, which confirms the robustness of our analysis and the conclusions obtained using the 7\,937 pixels detected in the three axes.

\section{Survival analysis of the resolved scaling relations}\label{sec:app_survival_analysis}
\begin{figure*}
    \centering
    \includegraphics[width=\textwidth]{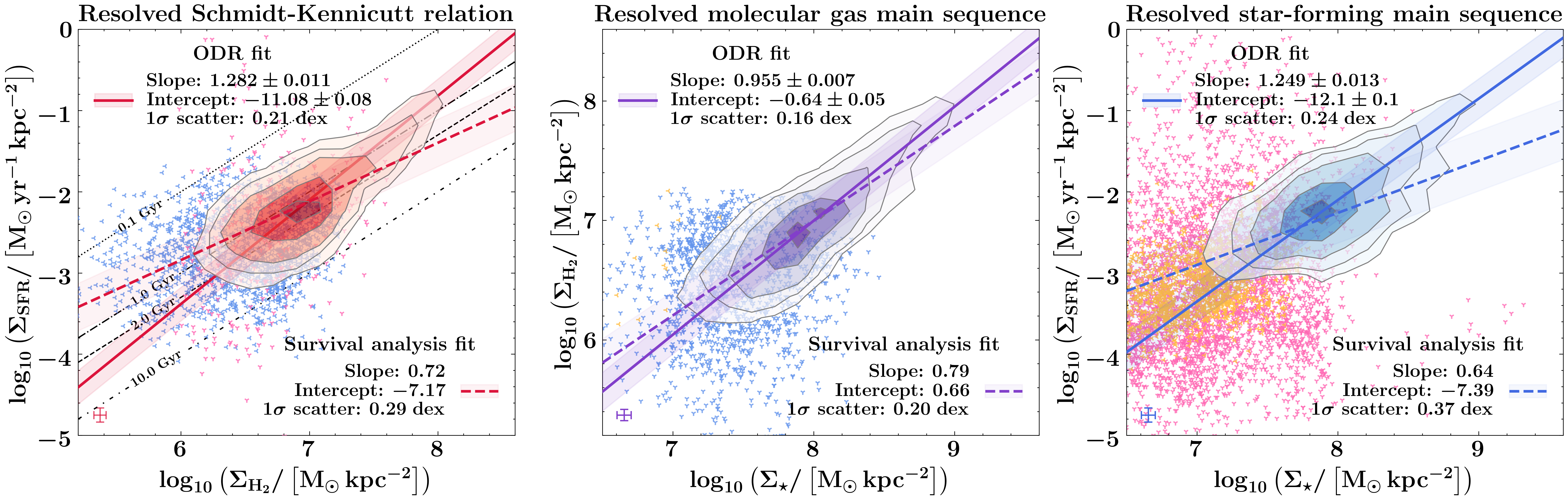}
    \caption{Resolved SRs for the 41 ALMA CO-CAVITY galaxies including upper limits and uncertainties in the fitting process. Lines and contours are similar than that of \cref{fig:triple_general_plot_extended}. In each panel, we show the upper limits: blue markers are $\Sigma_\mathrm{H_2}$ upper limits, pink markers are $\Sigma_\mathrm{SFR}$ upper limits, and yellow markers are $\Sigma_\star$ upper limits. For each relation we show the fit applied on the detected and censored values with the dashed lines, and the parameters for the fit in the lower-right corner of the panel, and we show the ODR fit with solid lines and its parameters in the upper-left corner of the panels. The orthogonal scatter of the ODR fit and the vertical scatter of the survival analysis fit are shown as a shadowed region surrounding each curve.}
    \label{fig:triple_general_plot_ODR_SA}
\end{figure*}

It has been established that non-detected pixels can have relevant impact when performing measurements at high spatial resolution \cite[e.g.][]{Pessa2021}. To account for this effect, we reproduce the same analysis as above using survival analysis and including upper limits in the three quantities we are studying, in opposition to the detected values characterised in \cref{sec:selection_det_spax}. These upper limits are values that were non-detected in one axis but detected in the other, within the elliptical disc delimited by the $d_{25}$ diameter. We apply linear fits to the data including or not these upper limits as censored data, following the methodology depicted by \cite{Wong2024}, and using the fitting pipeline \texttt{LEO-Py}~\citep{LEO_Py}. In \cref{fig:triple_general_plot_ODR_SA} we show how the three relations change when including the upper limits, and compare with the ODR fits reported in \cref{tab:rSRs_all_fits_full_sample_surveys}.

We obtain different results when comparing the parameters for the fits including the upper limits or excluding them, especially with the rSK, where the slope goes from $1.282\pm0.011$ to $0.72$ with upper limits. We also observe a larger scatter of 0.29\,dex, due to the larger uncertainties of the non-detected values. These variations and the large scatter in the relation show the extreme sensibility of the SRs to the pixel selection.

For the rMGMS, we observe closer results between both fits. This is easily explained by the number of non-detected values in each case and, most of all, the scatter of the non-detected values in the $\Sigma_\mathrm{SFR}$ axis: the rSK presents 1\,865 non-detected pixels spanning 3.9 and 9.0\,dex on the horizontal and vertical axes, respectively; whereas the rMGMS only has 1\,471 non-detected pixels spanning 2.7\,dex in the horizontal axis and 3.3\,dex along the vertical axis; and the rSFMS has 4\,009 non-detected pixels spanning 3.8 and 11.5\,dex, respectively. This is due to the closer sensitivity between $\Sigma_\mathrm{H_2}$ and $\Sigma_\star$: there are more non-detected values in the $\Sigma_\mathrm{H_2}$ and $\Sigma_\star$ maps that coincide with detected $\Sigma_\mathrm{SFR}$ values, whereas the same non-detected $\Sigma_\mathrm{H_2}$ pixels coincide more with non-detected $\Sigma_\star$ pixels.

The rSFMS also presents a shallower slope when considering the upper limits. In this case, there are more upper limits, due to the area covered by either of the two optical maps, which is generally larger than the area covered by the $\Sigma_\mathrm{H_2}$ map. We see that the two linear fits agree very well despite the presence of upper limits.

\cite{Wong2024} applied the same analysis and found closer results between their OLS fit and \texttt{LEO-Py} fit, due to the fact that both work using an OLS procedure, whereas we compare here with the ODR fit. Also, their upper limits are mainly affecting the rMGMS, whereas they are affecting all the relations in our work. Furthermore, the scatter values they calculate are showing a tighter rSK than rMGMS, which differs from our results and that of ALMaQUEST XI~\citep{Ellison2024}; but as mentioned in \cref{sec:full_sample_rSRs}, these discrepancies are hard to interpret since their data and that of ALMA CO-CAVITY are significantly different.

All the fits made including upper limits present larger $1\sigma$ scatter than the ODR fit. This is because \texttt{LEO-Py} is using an ordinary least-square fit, taking into account only the vertical residuals, whereas the $1\sigma$ scatter from the ODR fits is estimated taking into account the vertical and horizontal residuals. For this reason, we cannot compare both scatters directly. On the other hand, all the three slopes have decreased, getting to sub-linear values. Nonetheless, the rMGMS presents the smallest scatter of the three relations, even including the upper limits, and that the rSK is the second relation with the smallest scatter, followed finally by the rSFMS. This result is similar than what we find without including the upper limits in the analysis, showing that our initial analysis is robust.

\section{Linear-fit parameters for resolved scaling relations}\label{sec:app_parameters_linear_fits}
In this appendix we present the values of the parameters obtained using linear-fitting techniques for each of the galaxies in the ALMA CO-CAVITY sample.

In \cref{tab:rSK_fits_individual,tab:rMGMS_fits_individual,tab:rSFMS_fits_individual} we show the individual linear fits applied to the resolved SRs of individual galaxies. We applied two fits: an ODR fit\footnote{The ODR fit is made using Python's \texttt{odrpack 0.5.0} library (\url{https://pypi.org/project/odrpack/}).}, including only detected pixels in the two axes of the relation and without uncertainties, and a maximum likelihood (ML) fit on the detected values and upper limits, including also uncertainties, using \texttt{LEO-Py} pipeline. In some cases we do not present the results of the fits, either because the galaxy was not detected for this relation, or because the fit failed (when the slope is above 4.0 or the uncertainty of the slope or intercept is above 60\%). In the case of the rSFMS, we use all detected values in $\Sigma_\star$ and $\Sigma_\mathrm{SFR}$, even though some of these pixels correspond to non-detections in $\Sigma_\mathrm{H_2}$.

\begin{table}
\caption{Linear-fit parameters for the resolved Schmidt-Kennicutt relation (rSK).}
\label{tab:rSK_fits_individual}
\resizebox{\hsize}{!}{
    \begin{tabular}{c|ccc|ccc}
        \hline\hline
         ID & \multicolumn{3}{c|}{ODR fit (without upper limits)} & \multicolumn{3}{c}{ML fit (with upper limits)} \\
         & Slope & Intercept & Scatter & Slope & Intercept & Scatter \\
         &   &   & (dex) &   &   & (dex) \\
        \hline
        11248 & $1.05 \pm 0.07$ & $-9.3 \pm 0.4$ & $0.194$ & $0.697$ & $-6.972$ & $0.179$ \\
        26668 & $2.2 \pm 0.4$ & $-17 \pm 3$ & $0.208$ & $0.141$ & $-3.503$ & $0.202$ \\
        27289 & $0.92 \pm 0.09$ & $-8.6 \pm 0.6$ & $0.124$ & $0.698$ & $-7.080$ & $0.117$ \\
        27516 & $0.97 \pm 0.03$ & $-8.96 \pm 0.18$ & $0.201$ & $1.124$ & $-10.115$ & $0.136$ \\
        27657 & $1.82 \pm 0.21$ & $-14.7 \pm 1.5$ & $0.114$ & $0.788$ & $-7.507$ & $0.147$ \\
        32522 & $-$ & $-$ & $-$ & $-$ & $-$ & $-$ \\
        32597 & $2.31 \pm 0.24$ & $-17.7 \pm 1.7$ & $0.156$ & $1.389$ & $-11.333$ & $0.268$ \\
        34439 & $-$ & $-$ & $-$ & $-$ & $-$ & $-$ \\
        34718 & $1.79 \pm 0.12$ & $-14.6 \pm 0.8$ & $0.249$ & $1.352$ & $-11.664$ & $0.222$ \\
        37424 & $-$ & $-$ & $-$ & $0.142$ & $-3.365$ & $0.272$ \\
        38659 & $1.12 \pm 0.06$ & $-9.9 \pm 0.4$ & $0.166$ & $1.011$ & $-9.227$ & $0.171$ \\
        40294 & $-$ & $-$ & $-$ & $-$ & $-$ & $-$ \\
        42443 & $-$ & $-$ & $-$ & $-$ & $-$ & $-$ \\
        42595 & $1.218 \pm 0.014$ & $-10.48 \pm 0.10$ & $0.097$ & $1.159$ & $-10.056$ & $0.114$ \\
        46746 & $-$ & $-$ & $-$ & $-$ & $-$ & $-$ \\
        48125 & $1.53 \pm 0.08$ & $-12.9 \pm 0.6$ & $0.174$ & $1.162$ & $-10.318$ & $0.208$ \\
        48131 & $-$ & $-$ & $-$ & $-$ & $-$ & $-$ \\
        49935 & $1.9 \pm 0.8$ & $-15 \pm 6$ & $0.175$ & $0.460$ & $-5.528$ & $0.182$ \\
        50031 & $1.7 \pm 0.3$ & $-13.6 \pm 1.9$ & $0.171$ & $0.941$ & $-8.454$ & $0.258$ \\
        51089 & $-$ & $-$ & $-$ & $0.778$ & $-7.380$ & $0.252$ \\
        52854 & $-$ & $-$ & $-$ & $0.253$ & $-3.595$ & $0.288$ \\
        53449 & $1.33 \pm 0.06$ & $-11.3 \pm 0.4$ & $0.169$ & $1.209$ & $-10.466$ & $0.174$ \\
        53609 & $1.47 \pm 0.11$ & $-12.9 \pm 0.8$ & $0.167$ & $1.061$ & $-9.942$ & $0.246$ \\
        54706 & $0.95 \pm 0.03$ & $-8.56 \pm 0.18$ & $0.117$ & $0.842$ & $-7.810$ & $0.109$ \\
        55131 & $0.77 \pm 0.08$ & $-7.3 \pm 0.5$ & $0.131$ & $0.482$ & $-5.355$ & $0.082$ \\
        55734 & $1.61 \pm 0.09$ & $-13.7 \pm 0.7$ & $0.166$ & $1.312$ & $-11.369$ & $0.201$ \\
        57508 & $2.2 \pm 0.7$ & $-17 \pm 5$ & $0.159$ & $0.941$ & $-8.626$ & $0.181$ \\
        58154 & $2.15 \pm 0.07$ & $-17.1 \pm 0.5$ & $0.213$ & $0.839$ & $-8.024$ & $0.346$ \\
        58740 & $2.1 \pm 0.3$ & $-16.5 \pm 1.8$ & $0.156$ & $0.897$ & $-8.123$ & $0.228$ \\
        58741 & $1.54 \pm 0.14$ & $-13.1 \pm 1.1$ & $0.215$ & $1.294$ & $-11.228$ & $0.269$ \\
        58855 & $1.89 \pm 0.15$ & $-15.8 \pm 1.1$ & $0.155$ & $1.479$ & $-12.721$ & $0.249$ \\
        59266 & $1.41 \pm 0.06$ & $-12.0 \pm 0.4$ & $0.178$ & $0.335$ & $-4.537$ & $0.194$ \\
        59764 & $0.66 \pm 0.06$ & $-7.0 \pm 0.4$ & $0.158$ & $0.416$ & $-5.311$ & $0.152$ \\
        59902 & $0.83 \pm 0.06$ & $-7.9 \pm 0.4$ & $0.149$ & $0.612$ & $-6.491$ & $0.141$ \\
        60871 & $1.20 \pm 0.06$ & $-10.6 \pm 0.4$ & $0.269$ & $1.083$ & $-9.787$ & $0.211$ \\
        62323 & $1.10 \pm 0.09$ & $-9.4 \pm 0.6$ & $0.125$ & $0.698$ & $-6.691$ & $0.130$ \\
        63263 & $0.66 \pm 0.12$ & $-6.4 \pm 0.8$ & $0.162$ & $0.686$ & $-6.564$ & $0.001$ \\
        65288 & $1.52 \pm 0.12$ & $-12.7 \pm 0.8$ & $0.196$ & $0.997$ & $-9.274$ & $0.267$ \\
        65716 & $0.87 \pm 0.07$ & $-8.0 \pm 0.5$ & $0.193$ & $0.670$ & $-6.713$ & $0.001$ \\
        65783 & $1.02 \pm 0.03$ & $-8.77 \pm 0.22$ & $0.147$ & $0.986$ & $-8.513$ & $0.161$ \\
        65887 & $1.47 \pm 0.12$ & $-12.1 \pm 0.8$ & $0.173$ & $1.101$ & $-9.707$ & $0.149$ \\ \hline
    \end{tabular}}
    \tablefoot{
    For each detected galaxy, we show the slope, intercept, and scatter of the linear fit estimated using ODR regression procedure (without considering upper limits) and using \texttt{LEO-Py} ML fit (considering upper limits). Cells with no values and `$-$' symbol are either non-detected galaxies that were purposely not fitted (since they do not have any detected values), or galaxies where the ODR fit failed (due to an uncertainty higher than 60\% in the parameters for the fit, or because the slope was above 4.0).
    }
\end{table}

\begin{table}
    \caption{Linear-fit parameters for the resolved molecular gas main sequence (rMGMS) relation. Columns are the same as in \cref{tab:rSK_fits_individual}.}
    \label{tab:rMGMS_fits_individual}
    \resizebox{\hsize}{!}{
    \begin{tabular}{c|ccc|ccc}
        \hline \hline
        ID & \multicolumn{3}{c|}{ODR fit (without upper limits)} & \multicolumn{3}{c}{ML fit (with upper limits)} \\
         & Slope & Intercept & Scatter & Slope & Intercept & Scatter \\
         &   &   & (dex) &   &   & (dex) \\
        \hline
        11248 & $0.78 \pm 0.04$ & $0.7 \pm 0.3$ & $0.159$ & $0.656$ & $1.654$ & $0.173$ \\
        26668 & $0.44 \pm 0.07$ & $3.4 \pm 0.6$ & $0.200$ & $0.208$ & $5.169$ & $0.185$ \\
        27289 & $0.74 \pm 0.06$ & $1.1 \pm 0.4$ & $0.109$ & $0.667$ & $1.659$ & $0.097$ \\
        27516 & $-$ & $-$ & $-$ & $0.735$ & $1.044$ & $0.185$ \\
        27657 & $0.78 \pm 0.07$ & $1.0 \pm 0.5$ & $0.090$ & $0.618$ & $2.191$ & $0.058$ \\
        32522 & $-$ & $-$ & $-$ & $-$ & $-$ & $-$ \\
        32597 & $0.45 \pm 0.05$ & $3.4 \pm 0.4$ & $0.172$ & $0.386$ & $3.878$ & $0.154$ \\
        34439 & $-$ & $-$ & $-$ & $-$ & $-$ & $-$ \\
        34718 & $1.15 \pm 0.04$ & $-2.0 \pm 0.3$ & $0.132$ & $0.968$ & $-0.645$ & $0.131$ \\
        37424 & $0.50 \pm 0.04$ & $2.8 \pm 0.3$ & $0.172$ & $0.304$ & $4.272$ & $0.158$ \\
        38659 & $1.13 \pm 0.04$ & $-2.2 \pm 0.3$ & $0.116$ & $1.036$ & $-1.479$ & $0.068$ \\
        40294 & $-$ & $-$ & $-$ & $-$ & $-$ & $-$ \\
        42443 & $-$ & $-$ & $-$ & $-$ & $-$ & $-$ \\
        42595 & $1.139 \pm 0.014$ & $-2.09 \pm 0.11$ & $0.089$ & $1.064$ & $-1.473$ & $0.116$ \\
        46746 & $-$ & $-$ & $-$ & $-$ & $-$ & $-$ \\
        48125 & $0.81 \pm 0.04$ & $0.6 \pm 0.3$ & $0.152$ & $0.669$ & $1.760$ & $0.170$ \\
        48131 & $-$ & $-$ & $-$ & $-$ & $-$ & $-$ \\
        49935 & $-$ & $-$ & $-$ & $0.560$ & $2.479$ & $0.148$ \\
        50031 & $0.60 \pm 0.09$ & $2.0 \pm 0.7$ & $0.159$ & $0.493$ & $2.880$ & $0.167$ \\
        51089 & $0.54 \pm 0.05$ & $2.6 \pm 0.4$ & $0.121$ & $0.331$ & $4.175$ & $0.092$ \\
        52854 & $-$ & $-$ & $-$ & $0.052$ & $6.607$ & $0.149$ \\
        53449 & $0.81 \pm 0.04$ & $0.7 \pm 0.3$ & $0.175$ & $0.699$ & $1.539$ & $0.204$ \\
        53609 & $1.01 \pm 0.06$ & $-1.3 \pm 0.5$ & $0.120$ & $0.875$ & $-0.161$ & $0.155$ \\
        54706 & $-$ & $-$ & $-$ & $0.659$ & $1.840$ & $0.238$ \\
        55131 & $-$ & $-$ & $-$ & $0.763$ & $0.754$ & $0.070$ \\
        55734 & $0.646 \pm 0.023$ & $2.22 \pm 0.19$ & $0.152$ & $0.609$ & $2.525$ & $0.172$ \\
        57508 & $2.2 \pm 0.8$ & $-11 \pm 6$ & $0.115$ & $0.654$ & $1.629$ & $0.154$ \\
        58154 & $1.164 \pm 0.021$ & $-2.34 \pm 0.17$ & $0.114$ & $0.914$ & $-0.353$ & $0.149$ \\
        58740 & $0.41 \pm 0.08$ & $3.7 \pm 0.6$ & $0.182$ & $0.240$ & $5.019$ & $0.154$ \\
        58741 & $-$ & $-$ & $-$ & $0.842$ & $0.466$ & $0.243$ \\
        58855 & $-$ & $-$ & $-$ & $0.837$ & $0.289$ & $0.151$ \\
        59266 & $-$ & $-$ & $-$ & $0.623$ & $1.934$ & $0.133$ \\
        59764 & $0.69 \pm 0.04$ & $1.5 \pm 0.3$ & $0.149$ & $0.506$ & $2.916$ & $0.142$ \\
        59902 & $0.97 \pm 0.05$ & $-0.7 \pm 0.4$ & $0.124$ & $0.781$ & $0.703$ & $0.111$ \\
        60871 & $0.95 \pm 0.03$ & $-0.42 \pm 0.23$ & $0.164$ & $0.874$ & $0.201$ & $0.184$ \\
        62323 & $0.62 \pm 0.04$ & $2.0 \pm 0.3$ & $0.108$ & $0.549$ & $2.521$ & $0.104$ \\
        63263 & $-$ & $-$ & $-$ & $0.874$ & $-0.147$ & $0.001$ \\
        65288 & $1.08 \pm 0.07$ & $-1.4 \pm 0.5$ & $0.134$ & $0.856$ & $0.233$ & $0.138$ \\
        65716 & $1.36 \pm 0.10$ & $-3.5 \pm 0.7$ & $0.165$ & $1.039$ & $-1.066$ & $0.001$ \\
        65783 & $0.94 \pm 0.03$ & $-0.46 \pm 0.22$ & $0.143$ & $0.917$ & $-0.231$ & $0.146$ \\
        65887 & $-$ & $-$ & $-$ & $0.605$ & $1.888$ & $0.162$ \\ \hline
    \end{tabular}}
\end{table}

\begin{table}
    \caption{Linear-fit parameters for the resolved star-forming main sequence (rSFMS) relation. Columns are the same as in \cref{tab:rSK_fits_individual}.}
    \label{tab:rSFMS_fits_individual}
    \resizebox{\hsize}{!}{
    \begin{tabular}{c|ccc|ccc}
        \hline \hline
        ID & \multicolumn{3}{c|}{ODR fit (without upper limits)} & \multicolumn{3}{c}{ML fit (with upper limits)} \\
         & Slope & Intercept & Scatter & Slope & Intercept & Scatter \\
         &   &   & (dex) &   &   & (dex) \\
        \hline
        11248 & $1.70 \pm 0.08$ & $-15.1 \pm 0.6$ & $0.383$ & $0.997$ & $-10.035$ & $0.306$ \\
        26668 & $1.59 \pm 0.07$ & $-14.2 \pm 0.5$ & $0.386$ & $0.099$ & $-3.733$ & $0.426$ \\
        27289 & $0.82 \pm 0.04$ & $-8.6 \pm 0.3$ & $0.222$ & $0.680$ & $-7.652$ & $0.219$ \\
        27516 & $1.44 \pm 0.05$ & $-14.0 \pm 0.4$ & $0.372$ & $1.106$ & $-11.349$ & $0.343$ \\
        27657 & $1.67 \pm 0.05$ & $-14.8 \pm 0.3$ & $0.431$ & $1.185$ & $-11.516$ & $0.419$ \\
        32522 & $1.99 \pm 0.19$ & $-17.1 \pm 1.4$ & $0.617$ & $0.751$ & $-8.292$ & $0.460$ \\
        32597 & $1.43 \pm 0.12$ & $-12.8 \pm 0.9$ & $0.423$ & $1.007$ & $-9.776$ & $0.326$ \\
        34439 & $2.0 \pm 0.3$ & $-17.3 \pm 1.9$ & $0.535$ & $0.585$ & $-7.629$ & $0.464$ \\
        34718 & $1.66 \pm 0.09$ & $-15.0 \pm 0.7$ & $0.449$ & $1.294$ & $-12.376$ & $0.250$ \\
        37424 & $2.46 \pm 0.08$ & $-20.8 \pm 0.6$ & $0.424$ & $1.109$ & $-11.194$ & $0.518$ \\
        38659 & $1.29 \pm 0.04$ & $-12.6 \pm 0.3$ & $0.160$ & $1.213$ & $-12.069$ & $0.120$ \\
        40294 & $0.030 \pm 0.017$ & $-3.21 \pm 0.12$ & $0.765$ & $0.024$ & $-3.263$ & $0.386$ \\
        42443 & $2.52 \pm 0.14$ & $-21.2 \pm 1.1$ & $0.321$ & $1.290$ & $-12.290$ & $0.303$ \\
        42595 & $1.365 \pm 0.020$ & $-12.81 \pm 0.16$ & $0.212$ & $1.278$ & $-12.145$ & $0.170$ \\
        46746 & $0.132 \pm 0.012$ & $-3.38 \pm 0.09$ & $0.599$ & $0.137$ & $-3.450$ & $0.360$ \\
        48125 & $1.49 \pm 0.07$ & $-13.8 \pm 0.5$ & $0.363$ & $1.192$ & $-11.587$ & $0.294$ \\
        48131 & $3.6 \pm 0.7$ & $-29 \pm 5$ & $0.565$ & $0.743$ & $-8.475$ & $0.509$ \\
        49935 & $2.32 \pm 0.13$ & $-18.8 \pm 0.9$ & $0.482$ & $0.925$ & $-9.572$ & $0.269$ \\
        50031 & $0.75 \pm 0.04$ & $-7.9 \pm 0.3$ & $0.439$ & $0.524$ & $-6.441$ & $0.332$ \\
        51089 & $1.57 \pm 0.04$ & $-14.2 \pm 0.3$ & $0.192$ & $1.330$ & $-12.435$ & $0.196$ \\
        52854 & $1.98 \pm 0.12$ & $-16.9 \pm 0.8$ & $0.563$ & $0.828$ & $-8.671$ & $0.515$ \\
        53449 & $1.21 \pm 0.07$ & $-11.5 \pm 0.5$ & $0.309$ & $1.159$ & $-11.156$ & $0.229$ \\
        53609 & $1.68 \pm 0.07$ & $-16.5 \pm 0.6$ & $0.229$ & $1.535$ & $-15.361$ & $0.256$ \\
        54706 & $0.955 \pm 0.023$ & $-9.43 \pm 0.17$ & $0.262$ & $0.779$ & $-8.144$ & $0.315$ \\
        55131 & $1.88 \pm 0.08$ & $-15.7 \pm 0.5$ & $0.432$ & $0.751$ & $-8.082$ & $0.347$ \\
        55734 & $1.37 \pm 0.06$ & $-12.8 \pm 0.5$ & $0.367$ & $1.337$ & $-12.749$ & $0.185$ \\
        57508 & $3.35 \pm 0.18$ & $-26.9 \pm 1.3$ & $0.375$ & $0.808$ & $-8.774$ & $0.448$ \\
        58154 & $0.298 \pm 0.010$ & $-4.80 \pm 0.07$ & $0.579$ & $0.229$ & $-4.281$ & $0.488$ \\
        58740 & $1.14 \pm 0.05$ & $-10.8 \pm 0.4$ & $0.271$ & $0.852$ & $-8.671$ & $0.362$ \\
        58741 & $1.59 \pm 0.08$ & $-14.9 \pm 0.7$ & $0.285$ & $1.505$ & $-14.209$ & $0.202$ \\
        58855 & $1.58 \pm 0.08$ & $-15.3 \pm 0.7$ & $0.265$ & $1.582$ & $-15.415$ & $0.385$ \\
        59266 & $2.17 \pm 0.06$ & $-19.3 \pm 0.4$ & $0.313$ & $0.710$ & $-8.011$ & $0.334$ \\
        59764 & $0.81 \pm 0.03$ & $-8.49 \pm 0.22$ & $0.400$ & $0.588$ & $-7.026$ & $0.249$ \\
        59902 & $0.86 \pm 0.04$ & $-8.81 \pm 0.24$ & $0.420$ & $0.610$ & $-7.177$ & $0.301$ \\
        60871 & $1.00 \pm 0.05$ & $-10.0 \pm 0.4$ & $0.402$ & $1.003$ & $-10.035$ & $0.318$ \\
        62323 & $0.76 \pm 0.03$ & $-7.93 \pm 0.21$ & $0.292$ & $0.618$ & $-6.871$ & $0.306$ \\
        63263 & $2.7 \pm 0.3$ & $-21.8 \pm 1.8$ & $0.509$ & $1.043$ & $-10.156$ & $0.313$ \\
        65288 & $1.77 \pm 0.09$ & $-15.8 \pm 0.7$ & $0.228$ & $1.253$ & $-12.033$ & $0.158$ \\
        65716 & $1.14 \pm 0.05$ & $-10.7 \pm 0.4$ & $0.230$ & $1.048$ & $-10.100$ & $0.001$ \\
        65783 & $1.01 \pm 0.04$ & $-9.6 \pm 0.3$ & $0.367$ & $0.971$ & $-9.349$ & $0.286$ \\
        65887 & $1.42 \pm 0.07$ & $-13.4 \pm 0.5$ & $0.215$ & $1.182$ & $-11.676$ & $0.230$ \\ \hline
    \end{tabular}}
    \tablefoot{
    In this case we also show the results for galaxies that were not detected in ALMA datasets, since the rSFMS is not requiring values from the $\Sigma_\mathrm{H_2}$ axis.
    }
\end{table}

In \cref{tab:ODR_fits_running_medians} we show the results of the ODR fits to the running medians obtained for the three resolved SRs (presented in \cref{fig:scaling_relations_running_medians_all_sample}). They differ from the parameters presented in \cref{tab:rSRs_all_fits_full_sample_surveys} due to the lower number of points fitted (only 20 for the running medians and 7\,937 for the whole distributions) and especially in the low-$\Sigma$ values, where the full distributions are larger, increasing the fitted slope. As mentioned in the main text, a linear fit of the first order does not perfectly depict the trend followed by the running medians. The parameters proposed here should therefore be used with care.

\begin{table}
\caption{Linear-fit parameters to the running medians of the resolved SRs for the ALMA CO-CAVITY sample of 35 detected galaxies.}
\label{tab:ODR_fits_running_medians}
\centering
    \begin{tabular}{cccc}
        \hline\hline
        \multicolumn{1}{c}{Relation} & Slope & Intercept & Scatter \\
                                     &       &           &  (dex)  \\  \hline
        rSK    & $1.05 \pm 0.06$      & $-9.4 \pm 0.4$ & $0.11$    \\
        rMGMS  & $0.82 \pm 0.04$      & $0.5 \pm 0.3$  & $0.08$    \\
        rSFMS  & $0.90 \pm 0.06$      & $-9.2 \pm 0.5$ & $0.13$    \\ \hline \hline
    \end{tabular}
\end{table}

\section{Categories of galaxies based on their resolved scaling relations}\label{sec:app_categories_galaxies_rSRs}
In \cref{tab:categories_rSRs_ids} we show the visual classification of the 35 detected galaxies of our sample based on their resolved SRs. We classified them following the categories described in \cref{sec:groups_rSRs}. We note that galaxies can appear in several categories.

\begin{table*}[h]
    \caption{IDs of the galaxies of the ALMA CO-CAVITY visually classified based on the shape of their resolved SRs with respect to the full-sample relations.}
    \label{tab:categories_rSRs_ids}
    \begin{tabular}{M{0.28\textwidth}|M{0.15\textwidth}|M{0.15\textwidth}|M{0.15\textwidth}|M{0.15\textwidth}}
        \hline \hline
        Following full-sample trend & Different trend & Several trends & Above the full-sample trend & Below the full-sample trend \\ \hline
        \multicolumn{5}{c}{\textit{rSK}}                             \\ \hline        
        11248, 27289, 27516, 27657, 34718, 38659, 42595, 48125, 49935, 50031, 53449, 54706, 55734, 57508, 58741, 59266, 59902, 60871, 65288, 65716, 65887 & 26668, 32597, 51089, 54706, 55131, 58855, 59764, 63263 & 26668, 37424, 52854, 58154, 59266 & 32597, 58740, 62323, 63263, 65783 & 53609, 58855 \\ \hline
        \multicolumn{5}{c}{\textit{rMGMS}}                             \\ \hline       
        11248, 27289, 27516, 32597, 34718, 42595, 48125, 49935, 53449, 54706, 55131, 57508, 58154, 58740, 58741, 58855, 59902, 60871, 62323, 65288, 65716, 65783, 65887 & 26668, 50031, 51089, 52854, 55734, 62323 & 26668, 37424, 59266, 59764 & 27657, 55734 & 38659, 50031, 53609, 63263 \\ \hline
        \multicolumn{5}{c}{\textit{rSFMS}}                             \\ \hline       
        11248, 26668, 27289, 27657, 34718, 37424, 42595, 48125, 49935, 51089, 52854, 53449, 54706, 55131, 55734, 57508, 58741, 59902, 60871, 63263, 65288, 65716, 65887 & 50031, 58855, 59764 & 58154, 59266 & 32597, 58740, 62323, 65783 & 27516, 38659, 53609, 58855 \\ \hline
    \end{tabular}
    \tablefoot{
    As the groups are not exclusive, it is possible to find a galaxy in more than one group.
    }
\end{table*}

\begin{figure*}
    \centering
    \includegraphics[width=\textwidth]{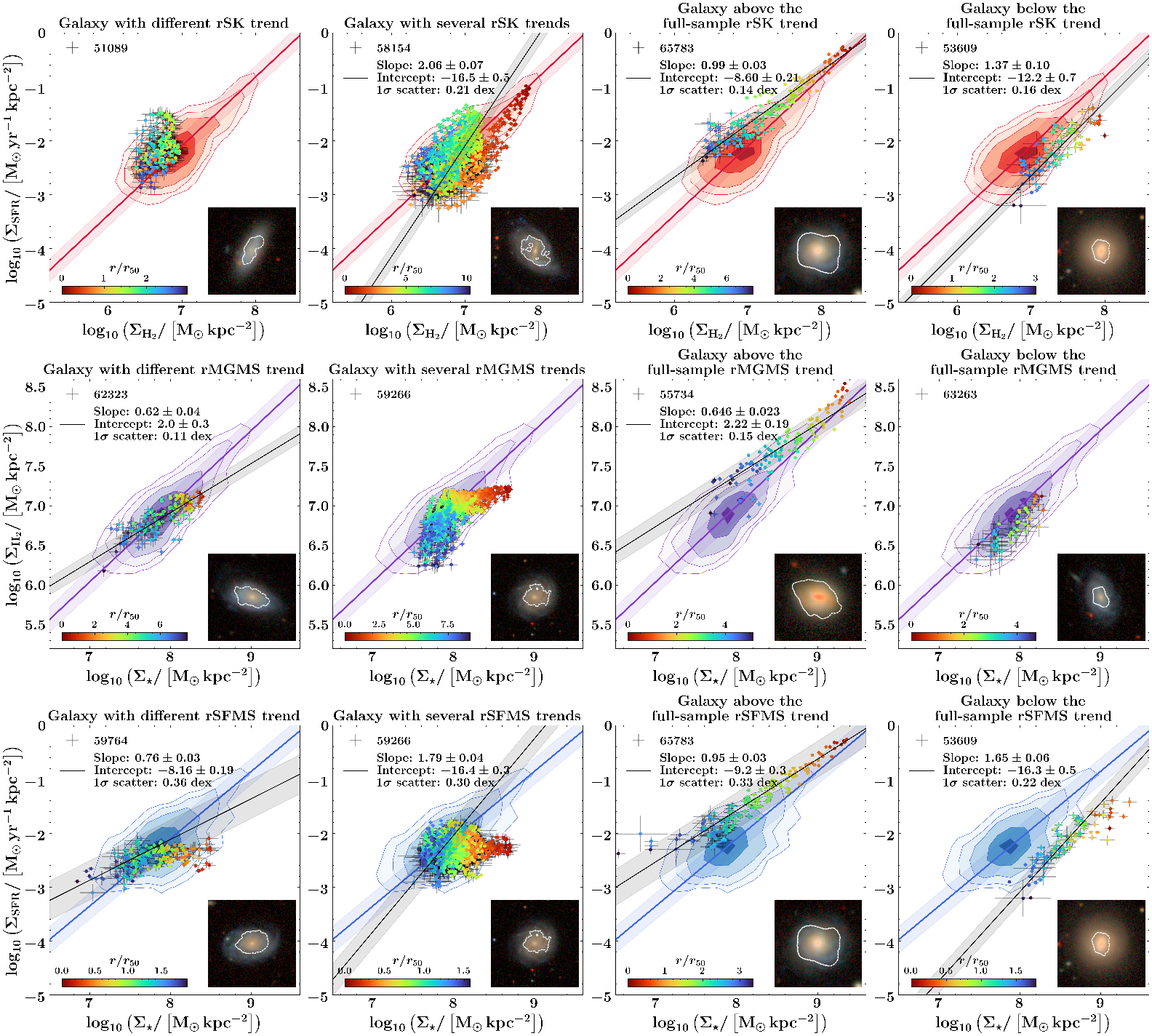}
     \caption{Resolved SRs of representative galaxies of the selected groups based on their resolved SRs: galaxies following a different trend (first column), galaxies with several trends (second column), galaxies above the trend traced by the full-sample distribution (third column), and below the trend (fourth column). For each panel, we show the distribution of the full sample of detected pixels in red (rSK, first row), purple (rMGMS, second row), and blue (rSFMS, third row). For each relation we show the ODR fit curve with the respective coloured solid line, and the $1\sigma$ scatter with the coloured shadow. The scatter coloured points show the individual pixels of each galaxy, with the ODR fit displayed with a solid black line (when the fit was successful), and the 1$\sigma$ scatter with the grey shadow. The parameters for each fit are shown in the upper-left legend. The colour of each data point indicates the galactocentric distance of the pixel and is labeled by the lower-left colour bar of each panel. In the lower-right corner of each panel we show an RGB image of the galaxy, with white contours showing the limits of the CO(1--0) mask.}
    \label{fig:resolved_scaling_relations_galaxy_categories_representative}
\end{figure*}

In \cref{fig:resolved_scaling_relations_galaxy_categories_representative} we show the representative galaxies for each of the four categories that do not follow the full-sample SRs (ii to v). Some galaxies only appear once in one of the categories, such as \texttt{62323} (first column, mid row in \cref{fig:resolved_scaling_relations_galaxy_categories_representative}), while others are in the same categories for several relations, such as \texttt{59266} (second column, mid and last rows), \texttt{65783} (third column, first and last rows), or \texttt{53609} (fourth column, first and last rows). This shows that the deviations from the typical SRs represented by the full sample can be produced by phenomena that are reflected in several or only one of the resolved SRs.

For instance, we can see in the atlas the cases of \texttt{53609} and \texttt{58154}, that we already discussed in the previous subsections. Data points from \texttt{53609} are found below the full-sample distribution in all three SRs, whereas \texttt{58154} is showing different trends in the rSK and rSFMS but no difference in the rMGMS with respect to the full-sample distribution. These differences could be explained by the morphology and internal properties of each galaxy. For instance, \texttt{53609} is a red galaxy found in the transition zone of the $M_\star-\mathrm{sSFR}$ diagram (Fig. 2, \citetalias{Espada2026}), its stellar component is dominated by red low-mass stars and its SFR is suppressed, resulting in low values of $\Sigma_\mathrm{SFR}$ and $\Sigma_\mathrm{H_2}$; meanwhile, \texttt{58154} has a bar and blue spiral arms, which result in a double trend in its rSK and rSFMS (as mentioned in \cref{sec:rSK_individual}).
    
\begin{figure*}
    \centering
    \includegraphics[width=\textwidth]{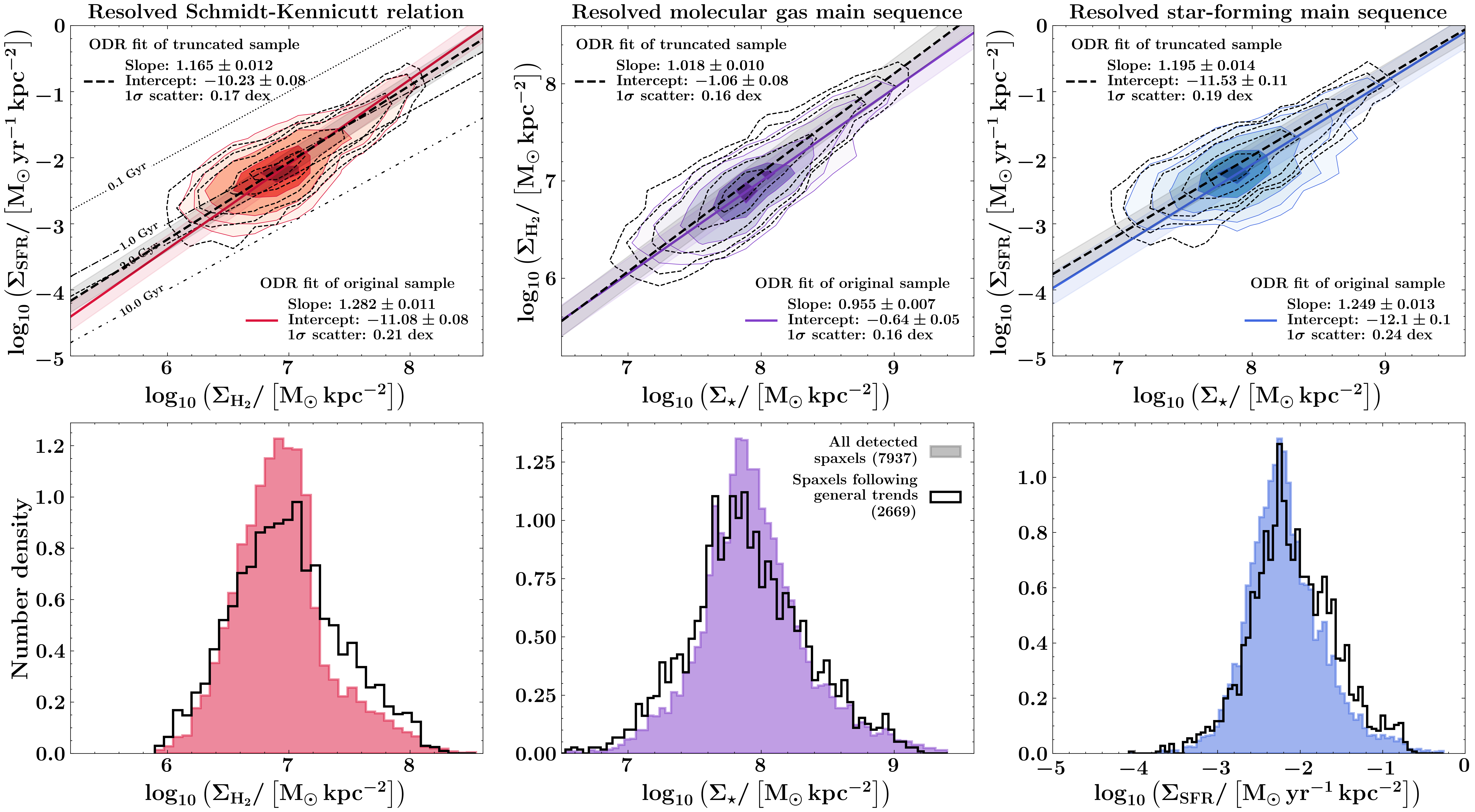}
    \caption{Resolved SRs (top panels) and surface density distributions (bottom panels) of the original detected sample of pixels and the truncated sample of galaxies. In the top panels we show with coloured contours the distribution of the original sample of 7\,937 pixels from the 35 detected galaxies, and with black-dashed contours the distribution of 2\,669 pixels from the 15 galaxies of the truncated sample. On top of each distribution we show the corresponding ODR fit, whose parameters are visible in the upper-left and lower-right corners of each panel. In the case of the rSK (left panel), we also show the equivalent slopes for depletion times at 0.1, 1.0, 2.0, and 10.0 Gyr, with black dotted and dashed lines.} In the bottom panels we show the distributions of $\Sigma_\mathrm{H_2}$ (left panel), $\Sigma_\star$ (mid panel), and $\Sigma_\mathrm{SFR}$ (right panel) of the original sample with the coloured histograms and the truncated sample with the black-hashed histograms. Note that the contours show levels of inclusion of pixels of 5, 10, 20, 50 and 80\%.
    \label{fig:resolved_scaling_relations_ODR_fits_distros_all_vs_truncated_sample}
\end{figure*}

When we remove the galaxies showing deviations from the full-sample trend of the resolved SRs, we obtain a subsample of 15 galaxies, with a total of 2\,669 pixels detected in the three axes. In \cref{fig:resolved_scaling_relations_ODR_fits_distros_all_vs_truncated_sample} we show the relations and surface density distributions of this subsample of pixels in comparison with the original sample of 7\,937 detected pixels. We see that the dynamic range remains unchanged in each case, as well as the fits to the resolved SRs.

\end{appendix}

\end{document}